\documentclass{elsart}
\usepackage{amsmath,amssymb,epsfig}
\newcommand {\bu} {{\boldsymbol{u}}}

\newcommand{\ap}{\alpha_+}
\newcommand{\am}{\alpha_-}
\newcommand{\bp}{\beta_+}
\newcommand{\bm}{\beta_-}

\newcommand{\eps}{\varepsilon}
\newcommand{\ab}{{\boldsymbol{\alpha}}}
\newcommand{\bb}{{\boldsymbol{\beta}}}
\newcommand{\fix}{\operatorname{\mathrm Fix}}

\newtheorem{theorem}{Theorem}
\newtheorem{corollary}{Corollary}

\begin{document}
\begin{frontmatter}
\title{Explicit centre manifold reduction and full bifurcation analysis
for an astigmatic Maxwell-Bloch laser}
\author[Southampton]{D.R.J.Chillingworth}
\author[Southampton]{G.D'Alessandro}
\author[Strathclyde]{F.Papoff}
\address[Southampton]{Department of Mathematics, University of
Southampton, \\
Southampton SO17 1BJ, England, UK}
\address[Strathclyde]{Department of Physics and Applied Physics, University of
Strathclyde, \\ 107 Rottenrow, Glasgow G4 0NG, Scotland, UK}


\begin{abstract}
We set up a general framework to study the bifurcations of nearly
degenerate modes of a Maxwell-Bloch laser model, and we apply it
specifically to the study of interactions of two modes of a laser with
broken circular symmetry. We use an explicit centre manifold reduction
to analyse the behaviour of this laser. Complex bifurcation sequences
involving single mode solutions, mode locking and mode beating regimes
are predicted.  These sequences are organized by a Hopf bifurcation
with $1:1$ resonance and a $Z_2$-symmetric Takens-Bogdanov
bifurcation.  Numerical simulations of the original system show good
agreement with theory.
\end{abstract}


\end{frontmatter}

\section {Introduction}

Symmetries pose considerable constraints on the configuration of
stationary solutions and on the time evolution of dynamical systems as
well as on their bifurcation behaviour
~\cite{sattinger1979,golubitsky1985,golubitsky1988,chossat00a}, to the
extent that the presence of symmetries can often dictate the existence
of certain solution branches independently of the specific model under
consideration. These ideas have been fruitfully used in optics either
to classify the patterns produced by experiments~\cite{green90a} or by
models~\cite{papoff99a,papoff99b,papoff00a}, and to infer from these
the symmetry of the systems under investigation.  They have also been
used to study in great detail the structure of the bifurcations of
normal form equations with the same symmetry as the
experiments~\cite{dangelo92a,lopezruiz93a,krauskopf00a}. Despite the
considerable success of these approaches, there are several
limitations. In the first case, the comparison between theory and
experiments is qualitative and the bifurcation diagram is not fully
explored. In the second case, there is no relation between the
parameters used in the normal form analysis and the physical
parameters and, furthermore, it is impossible to find where the normal
forms cease to be a good approximation of the system studied.  A
popular alternative is to ignore symmetry altogether and to study
models, usually nonlinear partial or integro-differential equations,
that are derived from first principles.  An advantage of this approach
is that such models can be valid both close to and far from the first
instability.  A disadvantage is that most of the analysis has to be
performed numerically, which means in practice that the bifurcation
diagram is rarely analysed completely.

In this paper we derive normal form equations for a laser model in the
presence of a typical breaking of cylindrical symmetry of the
cavity. We then study in detail the case in which the symmetry is
broken by astigmatism.  In this way we are able to reduce a partial
differential model to a system of ordinary differential equations and
to take full advantage of the symmetry in the study of the bifurcation
diagram without losing the physical meaning of the control
parameters. Furthermore, we are able to find the range of the control
parameters in which the normal form is in quantitative agreement with
the full model and where the agreement is only qualitative.  In the
example discussed in this paper, we show that the normal forms predict
the correct bifurcation structure and amplitude of the solutions for
small values of the control parameter; for larger values only the
nature of the bifurcation points is determined (see
Figure~\ref{fig:NumSim}).  The bifurcations turn out to be organised
by a $Z_2$-symmetric Takens-Bogdanov point and by a resonant Hopf
bifurcation.  A Takens-Bogdanov bifurcation
commonly occurs in this type of problem
(see~\cite{dacosta81a,rucklidge01a}, for example) and the bifurcation
diagrams in~\cite{dangelmayr91a} sketch some key features of
associated branching behaviour, but here we provide a complete picture
with the full state diagram (bifurcation set with associated phase
portraits) for the normal form of our equations for a wide range of
physically significant parameters.

The rest of this paper is organised as follows. In the next section we
introduce a specific laser model, namely the Maxwell-Bloch two-level
laser~\cite{lugiato85a}, and carry out an explicit centre manifold
reduction at the first bifurcation from the trivial solution as the
pump parameter is increased.  We then in Section~\ref{TwoModes} look
more closely at the particular case of astigmatic symmetry-breaking
from geometric $O(2)$ symmetry to $D_2$ symmetry.  The additional
$S^1$ phase-shift yields a bifurcation problem with broken $O(2)\times
S^1$ symmetry, much studied in the
literature~\cite{golubitsky1988,dangelmayr91a,golubitsky85a,golubitsky87a}.
We find, however, that the thorough analysis of responses to linear
symmetry-breaking perturbations given in~\cite{dangelmayr91a}, the
treatment most relevant to our problem, does not cover all important
cases as it assumes the equivalent of nonzero detuning
(see~Section~\ref{TwoModes}) that may not hold in this problem.  For
the zero detuning case we use the normal form for Hopf bifurcation
with $1:1$ resonance~\cite{takens74a}, and note here how a nonlinear
coordinate change converts the problem again into one with $O(2)\times
S^1$ symmetry, but now with the roles of $SO(2)$, the group of spatial
rotation, and $S^1$ interchanged.  This has interesting implications
for the interpretation of periodic solutions in terms of standing or
travelling waves.

Finally, we compare the results with numerical simulations for the
original problem.

\section{Laser normal forms in the presence of generic symmetry breaking}
\label{NormalForms}

We assume that the laser under study is a ring cavity gas laser with
approximate cylindrical symmetry as shown schematically in
Figure~\ref{fig:laser}.  This is roughly the same type of laser as
used in the experiments of Reference~\cite{green90a} and it has also
been used in many other
experimental~\cite{tredicce89a,dangoisse92a,louvergneaux96a,louvergneaux98a,louvergneaux98b,labate97a,ciofini98a,coates94a}
and
theoretical~\cite{lopezruiz93a,narducci86a,lugiato90a,solari90a,brambilla94a}
studies of pattern formation in lasers.  We use here a standard model
for such a system~\cite{lugiato90a}, incorporating a number of
assumptions.

First we assume that the light is linearly polarised and that the active
medium is isotropic, so that the electric field amplitude $\hat F$ can
be represented by a scalar field.

Secondly we assume that the active medium can be represented as a
collection of two-level atoms (Maxwell-Bloch two-level laser),
\emph{i.e.} that only two atomic levels are involved in the lasing process:
this allows us to use only two functions to represent the atomic
medium, namely the polarisation and the population inversion.  The
polarisation $\hat P$ represents the induced polarisation field of the
atoms, while the population inversion $N$ measures the difference in
population between the excited and the ground states.

Thirdly we assume that either the laser is producing a continuous wave
of light or, if it is pulsed, that the pulses are long compared to the
light period.  In this case we can write the polarisation and the
electric field as the product of a rapidly varying phase term and
a slowly varying amplitude:
\begin{eqnarray}
 & \hat F(x,y,z,t) = & \frac{1}{2} 
  \left [ F(x,y,z,t) e^{i(k_A z - \omega_A t)} + \mbox{c.c.} \right ] 
  \label{Ffast} \\
 & \hat P(x,y,z,t) = & \frac{1}{2} 
  \left [ P(x,y,z,t) e^{i(k_A z - \omega_A t)} + \mbox{c.c.} \right ] , 
  \label{Pfast}
\end{eqnarray}
where $\omega_A=ck_A$ is the frequency of the atomic transition, with
$c$ the speed of light and $k_A$ the corresponding wavenumber.  The
amplitudes $F$ and $P$ are assumed to be varying much more slowly than
$\omega_A$ (slowly varying approximation) so that the second
derivatives of $F$ and $P$ with respect to $z$ and $t$ in Maxwell's
wave equation can be neglected with respect to their first
derivatives~\cite{svelto1998}.

Finally, we assume that the gain and losses per cavity round trip are
small.  This allows us to neglect the variation of the electric field
amplitude $F$ and of the atomic variables $P$ and $N$ along the cavity
axis, so that the problem is reduced to two spatial dimensions
(\emph{i.e.} the centre plane of the active medium) and time.
\smallskip

Under these hypotheses, we can write the equations for the model
as~\cite{lugiato90a} 
\begin{eqnarray}
 & & \frac{\partial F}{\partial t} =
   {\mathcal L} F +
   P , \label{fpqeqn} \\
 & & \frac{\partial P}{\partial t} = -P + \chi F + F N , 
   \label{Peqn} \\
 & & \frac{\partial N}{\partial t} = - \gamma \left [ N + 
    \frac{1}{2} \left ( F \bar{P} + \bar{F} P \right ) \right ] , 
   \label{Neqn}
\end{eqnarray}
where the over-bar symbol indicates complex conjugate.  The linear
operator ${\mathcal L}$ is usually a partial differential operator which
depends upon the laser cavity and it is related to the propagation in
the empty cavity, \emph{i.e.} the cavity without the active
medium~\cite{lugiato90a}. In these equations time $t$ is
non-dimensional and is in units of the polarisation decay time.  For a
very large class of cavities, called {\it stable} in laser physics,
the electric field remains confined within the cavity. For these
cavities, the operator ${\mathcal L}$ has a system of eigenfunctions
$A_n(x,y)$
\begin{equation}
 {\mathcal L} A_{n}(x,y) =  \beta_n A_{n}(x,y), 
 \label{PropOp}
\end{equation}
which form an orthonormal basis for the space ${\mathcal H}$ of $L_2$ complex
functions defined in the transverse plane.  The lasers we consider
here are in this class.  The functions $A_n(x,y)$ are called {\it
empty cavity modes} because they are also eigenfunctions of the
operator ${\mathcal R_C}$ that propagates the field once around the empty
cavity from $z=0$ back to itself, namely
\begin{equation}
 {\mathcal R_C} A_{n}(x,y,0) = e^{\mu_{n}} A_{n}(x,y,0) .
 \label{CavModEqn}
\end{equation} 
The complex coefficient $\beta_n$ represents the attenuation and phase
shift of the mode $n$ per unit time.  It is related to the eigenvalue
of the mode given in~(\ref{CavModEqn}) by the relation $\beta_n =
\mu_n/T_c + i \hat \delta$, where $T_c$ is the cavity round trip time
and $\hat \delta$ is a detuning parameter,
\begin{equation}
 \hat \delta = \omega_A - \omega_C .
 \label{deltaDef}
\end{equation}
Here $\omega_C$ is a reference frequency with respect to which the
propagation phase shifts of the modes are measured. The case in which
${\mathcal L}$ is an integral operator can be dealt with in the same
way, but the previous relations are slightly different.

The energy fed into the laser is represented by the pump
parameter $\chi=\chi(x,y)$ and the decay rate of the population
inversion is given by $\gamma$.  In all that follows we assume that
$\chi$ is space-independent.  The appropriate calculations for a
space-dependent pump are sketched in Appendix~\ref{SpaceDepPump}.

Equations~(\ref{fpqeqn}-\ref{Neqn}) have a trivial solution $F=P=N=0$
that is stable for sufficiently small positive values of $\chi\,$: if
too little energy is provided the laser remains switched off.  The
linear stability of the trivial solution with respect to a small
perturbation $(\tilde F, \tilde P, \tilde N)$ is given by
\begin{equation}
\frac{\partial}{\partial t}
   \left ( 
     \begin{array}{c} \tilde F \\ \tilde P \\ \tilde N \end{array}
   \right )=
   \left ( 
     \begin{array}{ccc} 
     {\mathcal L} & 1 & 0 \\
     \chi & -1 &  0 \\
     0 & 0 & -\gamma \end{array}
   \right )
   \left ( 
     \begin{array}{c} \tilde F \\ \tilde P \\ \tilde N \end{array}
   \right )  \label{fpNeqnLin}
\end{equation}
and depends also on the perturbation itself. However, one can see that
$\tilde N$ is decoupled from $\tilde F, \tilde P$ and it is always
damped. It is convenient to expand both $\tilde F$ and $\tilde P$ in
terms of the eigenfunctions $A_n$ of ${\mathcal L}$. By projecting the
equations for $\tilde F$ and $\tilde P$ onto the $A_n$, we obtain an
infinite block-diagonal matrix with blocks of dimension two. The
eigenvalues of these blocks are given by
\begin{equation}
 \lambda_{n} = \frac{1}{2}
  \left [
    -1 + \beta_{n} \pm \sqrt{(1 + \beta_{n})^2 + 4 \chi}
  \right ]
 \label{lambda_n}
\end{equation}
and the corresponding eigenvectors are 
\begin{equation}
 \bu_n = \frac{1}{{\mathcal N}_n}
   \left ( 
     \begin{array}{c} 1 + \lambda_n \\ \chi \\ 0 \end{array}
   \right )
   A_n(x,y) 
 \label{bu_n}
\end{equation}
where ${\mathcal N}_n$ is a normalisation factor  chosen so that 
$(\boldsymbol{v}_n,\bu_n)=1$.  Here $\boldsymbol{v}_n$ is the
corresponding eigenvector of the adjoint matrix,
\begin{equation}
 \boldsymbol{v}_n = \frac{1}{\bar{\mathcal N}_n}
   \left ( 
     \begin{array}{c} 1 + \bar{\lambda}_n \\ 1 \\ 0 \end{array}
   \right )
   A_n(x,y) \, ,
\end{equation}
and the inner product is defined by
\begin{equation}
 (\boldsymbol{v}, \bu  ) = \int \bar{\boldsymbol{v}}^T\bu \, d x d y .
\end{equation}
We have, therefore, that
\begin{equation}
 {\mathcal N}_n^2 = (1 + \lambda_n)^2 +   \chi  .
\end{equation}
From equation~(\ref{lambda_n}) we can calculate that the trivial solution
loses its stability through a Hopf bifurcation of frequency
\begin{equation}
 \omega_{\hat n} = - \frac{\Im(\beta_{\hat n})}{1 - \Re(\beta_{\hat n})} 
 \label{omegaHopf}
\end{equation}
when $\chi = \chi_0(\hat n)$, where 
\begin{equation}
 \chi_0(n) = - \Re(\beta_n)
  \left ( 1 + \omega_n^2 \right ) .
 \label{ThreshFunct}
\end{equation}
The index $\hat n$ is defined as an element of the set ${\mathcal J}
\subset {\mathbb N}$ of those values of $n$ that minimise $\chi_0(n)$
over all the integers.

In the presence of unperturbed symmetry there may be degenerate
eigenvalues.  In other words, the set ${\mathcal J}$ has more than one
element.  Eigenvectors corresponding to eigenvalues $\lambda_n,
n\in{\mathcal J}$ are called {\it active modes}.  In general, we can
assume that the number $M$ of elements of ${\mathcal J}$ is finite, in
which case we say that the unperturbed system has degeneracy $M$: all
the active modes have eigenvalue with zero real part and imaginary
part given by~(\ref{omegaHopf}).  All the other modes have eigenvalue
with negative real part.  If the system is slightly perturbed the
degeneracy is either partially or totally lifted and the eigenvalues
of the active modes are no longer all equal.  However, we assume that
the perturbation is sufficiently small and that $\chi$ is sufficiently
close to $\chi_0(\hat n)$ to ensure that
\begin{equation}
 \Re(\lambda_n) < 0 \qquad \forall n \notin {\mathcal J}.
 \label{validCM}
\end{equation}
In other words, the modes that do not belong to the set ${\mathcal J}$ are
still linearly damped.  We can separate the field $\bu=(F,P,N)^T$ into
components along the space spanned by the active modes and its
orthogonal complement:
\begin{equation}
 \bu = \sum_{n \in {\mathcal J}} f_n \bu_n + \bu_\perp ,
 \label{SplittingU}
\end{equation}
where $\bu_\perp \equiv (F_\perp,P_\perp,N)^T$ is defined by
\begin{equation}
 (\boldsymbol{v}_n, \bu_\perp) = 0 , \hspace{20mm} \forall n \in {\mathcal J} .
 \label{OrthoRelUPerp}
\end{equation}
The equations for the dynamics of the active modes can be obtained by
a centre manifold reduction~\cite{carr81a} of
equations~(\ref{fpqeqn}-\ref{Neqn}). The nonlinearities in these
latter equations are quadratic. As a consequence, the centre manifold
reduction is greatly simplified and the equations for the active modes
can be obtained up to third order by ``simply'' projecting the
equations~(\ref{fpqeqn}-\ref{Neqn}) onto the active modes:
\begin{equation}
 \frac{d}{d t} f_n = \left ( \boldsymbol{v}_n, \frac{d \bu}{d t} \right ) ,
  \hspace{20mm} n \in {\mathcal J} .
 \label{EqActModesFull}
\end{equation}
After rather lengthy calculations (detailed in
Appendix~\ref{DetailedCalc}) we obtain that the amplitudes of the
active modes are given by
\begin{equation}
 \displaystyle \frac{d}{d t} f_n = \lambda_n f_n - 
   \chi \sum_{p,j,k \in {\mathcal J}} B_{n p j k} \,
        (A_n, A_j \bar A_k A_p) f_j \bar f_k f_p ,
   \, n \in {\mathcal J} \label{ActModEqn} 
\end{equation}
where
\begin{equation}
 B_{n p j k} = \frac{(1+\lambda_p) [1 + (\lambda_j + \bar \lambda_k)/2]}
        {[ 1 + (\lambda_j + \bar \lambda_k)/\gamma ] 
         {\mathcal N}_n {\mathcal N}_p 
         {\mathcal N}_j \bar {\mathcal N}_k} \, .
\end{equation}

\section{Two modes and $D_2$ geometrical symmetry: an example}
\label{TwoModes}

\subsection{Introduction}

Equation~(\ref{ActModEqn}) is very general: it can be used to describe
the bifurcation structure of a mean field limit two level
Maxwell-Bloch laser at an eigenvalue of arbitrary (finite)
multiplicity.  The methods used in deriving it are also completely
general. However, the great advantage of using a two level
Maxwell-Bloch laser model is that the nonlinear terms are relatively
simple and hence the centre manifold reduction of the laser equations
contains fairly few and manageable terms.

In order to illustrate the use of equation~(\ref{ActModEqn}) we
consider in this section a particularly simple case: we assume that
the perfectly symmetric laser has two degenerate modes at threshold
and that symmetry breaking, for example as caused by astigmatism,
reduces the geometrical symmetry from $O(2)$ to $D_2$ and lifts the
degeneracy by changing the mode resonance frequency.  The geometrical
symmetry of the astigmatic cavity is combined with the phase
invariance $S^1$- symmetry of the laser equations to give the symmetry
group $Z_2 \times S^1$ (see below).  This problem has been studied in
depth by many
authors~\cite{dangelmayr91a,golubitsky85a,golubitsky87a}.  Dangelmayr
and Knobloch~\cite{dangelmayr91a} have provided a fairly comprehensive
analysis of all the possible bifurcation scenarios induced by the
symmetry breaking from $O(2)\times S^1$ to $Z_2\times S^1$ in a
generic system in normal form. In the laser context, various
authors~\cite{dangelo92a,lopezruiz93a,solari90a,lopezruiz94a,boscolo97a,degen00a}
have studied the effect of symmetry breaking on mode interactions for
simple laser modes.  In a similar vein,
Vladimirov~\cite{vladimirov98a} studies the interaction between
longitudinal modes in a bidirectional ring laser.

This section contributes to these studies by highlighting the fact
that the two interacting modes need not be standard Gauss-Hermite
modes but can be any cavity mode: for example they could be the modes
of a waveguide cavity as studied
in~\cite{papoff99a,papoff99b,papoff00a}.  Moreover, we include in our
analysis a bifurcation scenario with unexpected additional symmetry
that is excluded from the generic studies in~\cite{dangelmayr91a} and
which arises when the frequency $\omega_C$ of the unperturbed cavity
mode and the frequency $\omega_A$ of atomic transmission coincide. We
now explain this in more detail.

Perturbations that break the symmetry typically cause the two modes to
have slightly different frequencies.  By a suitable choice of
reference frequency we can write their coefficients $\beta_n$ as
\begin{equation}
 \beta_n = \kappa \left \{ -1 + i 
    \left [ \delta + (-1)^n \eta \right ] \right \} \hspace{20mm}
    n = 1,2
 \label{betaExample}
\end{equation}
where $\kappa$ is the mode amplitude decay rate and $\delta = \hat
\delta / \kappa$ is the scaled cavity-atom detuning parameter.  We
have relabelled the modes so that the two active modes have indices
$1$ and $2$ and we have changed the reference frequency so that the
active modes have frequencies $\kappa(\delta\pm\eta)$.  In this
notation $\eta$ plays the role of symmetry breaking parameter: if
$\eta=0$ the two modes are degenerate and the system undergoes a
standard $O(2) \times S^1$ symmetric Hopf bifurcation.  If
$\delta=\pm\eta$ then atomic transition is resonant with mode $n=1$
and $n=2$ respectively.  The mode with frequency closest to the atomic
frequency (so that $\Im(\beta_n)$ is closest to zero) is the first to
become unstable as the pump parameter is increased from below to above
threshold.  However, if the atomic frequency is exactly in between
those of the two modes, \emph{i.e.} if $\delta=0$, then the laser has
no criterion by which to choose the most unstable mode and like
Buridan's ass does not know what to do.  Unlike the unfortunate
animal, though, it does not die of hunger but exhibits dynamical
behaviour that is an interesting ``combination'' of features of the
dynamics observed for $\delta$ small and nonzero.

\subsection{Centre Manifold Equations}

We write the pump parameter $\chi$ as
\begin{equation}
 \chi = \kappa ( 1 + \varepsilon ) .
\end{equation}
From~(\ref{betaExample}) and~(\ref{lambda_n}) we see that at resonance
$\delta = \pm \eta$ the threshold value of the pump parameter is $\chi
= \kappa$, that is $\varepsilon = 0$.  The assumption of being close
to threshold therefore becomes that $\varepsilon$ is a small
parameter.  Moreover, we assume also $| \delta \pm \eta| \ll 1$.  To
write the equations for the two modal amplitudes we
simplify~(\ref{ActModEqn}) by first expanding the coefficient of the
linear term up to first order in $\varepsilon$ and up to second order
in $\delta \pm \eta\,$:
\begin{equation}
 \lambda_n = \mu \varepsilon - \mu^3 [ \delta + (-1)^n \eta ]^2 +
             i \mu [ \delta + (-1)^n \eta ] + \cdots
\end{equation}
where
\begin{equation}
 \mu \equiv \frac{\kappa}{1 + \kappa} .
\end{equation}
This expansion ensures that we keep track of all the physically
relevant features of the model: in particular, the second order term
in $\delta \pm \eta$ is responsible for making the threshold value of
the pump parameter dependent on the detuning.  We also expand the
coefficients of the nonlinear terms, but only up to order zero.  In
this limit equation~(\ref{ActModEqn}) reduces to
\begin{eqnarray}
 & \displaystyle \frac{d}{d t} f_n = & \mu \left \{
   \varepsilon - \mu^2 
     \left [ \delta + (-1)^n \eta \right ]^2 + 
  i [ \delta + (-1)^n \eta ] \right \} f_n \nonumber \\
 & & 
  - \frac{\mu}{1+\kappa} \sum_{j,k,p=1}^2 (A_n, A_j \bar A_k A_p)
   f_j \bar f_k f_p , \hspace{15mm} n=1,2\,.
 \label{AmplEqnTwoModes1}
\end{eqnarray}
In order to determine which of the nonlinear terms 
of~(\ref{AmplEqnTwoModes1}) are nonzero we use the
fact that the geometrical symmetry of the unperturbed system is $O(2)$
and that the perturbation is small so that we can compute the
projection integrals $(A_n,A_j\bar A_kA_p)$ assuming that the
modes $A_n(x,y)$ are the modes of the system with full symmetry.  

The (complex) 2-dimensional eigenspace for the linearisation of
(\ref{AmplEqnTwoModes1}) at the first threshold, \emph{i.e.} at the
bifurcation point where the zero field solution loses its stability,
must be a subspace of some particular isotypic component of the action
of $O(2)$ on the $L_2$ function space $\mathcal H$ spanned by the
cavity modes (see \cite[Theorem XII, 3.5]{golubitsky1988} for
example), and we can therefore assume that the modes $A_n(x,y)$,
represented in polar coordinates $(r,\psi)$ as $A_n(r,\psi)$, can be
written as (complex) linear combinations of $\cos(m\psi)$ and
$\sin(m\psi)$ for some $m \in {\mathbb N}$.  If we suppose that the
$x$ and $y$ axis are aligned with the symmetry axis of the perturbed
system then we may take
\begin{equation}
 A_1(r,\psi) = A_0(r) \cos(m \psi) 
 \quad \mbox{and} \quad 
 A_2(r,\psi) = A_0(r) \sin(m \psi) ,
 \label{sincosmodes}
\end{equation}
where $A_0(r)$ represents the radial profile of the two modes. 
The projection  integrals in~(\ref{AmplEqnTwoModes1}) have
values
\begin{equation}
 (A_n, A_j \bar A_k A_p) = {\mathcal M} 
  \left \{ 
   \begin{array}{ll@{\hspace{10mm}}l}
    3 \pi / 4 & & n=j=k=p \\*[4mm]
    \pi / 4 & & \parbox[c]{40mm}{\baselineskip 10pt
     $n=j \ne k=p$ \\
     {\small [all permutations of the \hfill \\ 
              four indices are allowed]}} \\*[7mm]
    0 & & \mbox{otherwise} 
   \end{array}
  \right .
 \label{SelRules}
\end{equation}
where the constant ${\mathcal M}$ is given by
\begin{equation}
 {\mathcal M} = \int_0^\infty A_0^2(r) \bar A_0^2(r) r \, d r .
 \label{defcalM}
\end{equation}
Using~(\ref{SelRules}) we can write~(\ref{AmplEqnTwoModes1}) as
\begin{eqnarray}
 & \displaystyle \frac{d}{d t} f_n = & \mu \left \{
   \varepsilon - \mu^2 
     \left [ \delta + (-1)^n \eta \right ]^2 + 
  i [ \delta + (-1)^n \eta ] \right \} f_n 
   \label{AmplEqnTwoModes2}\\*[3mm]
 & &  
  - \frac{\mu \pi {\mathcal M}}{4 (1+\kappa)} 
   \left [ ( 3 |f_n|^2 + 2 |f_p|^2 ) f_n + \bar f_n f_p^2 \right ] ,
   \hspace{5mm} n=1,2 , \, \, \,  p = 3 - n. \nonumber 
\end{eqnarray}
To simplify the notation we scale time and the modal amplitudes by
defining
\begin{equation}
 \tau = \mu t , \hspace{15mm}
 \alpha_{+} = \sqrt{\frac{ \pi {\mathcal M}}{4 (1+\kappa)}} f_1 , \hspace{15mm}
 \alpha_{-} = \sqrt{\frac{ \pi {\mathcal M}}{4 (1+\kappa)}} f_2 , 
\end{equation}
so that we can rewrite~(\ref{AmplEqnTwoModes2}) as
\begin{eqnarray}
 & \hspace{-5mm} \dot \alpha_+ = & \left [
   \varepsilon - \mu^2 \left ( \delta - \eta \right )^2 + 
  i ( \delta - \eta ) \right ] \alpha_+ - \left ( 3 |\alpha_+|^2 + 2 |\alpha_-|^2 \right ) \alpha_+ -
     \bar \alpha_+ \alpha_-^2 \, ,  \label{alphaPlus} \\*[3mm]
 & \hspace{-5mm} \dot \alpha_- = & \left [
   \varepsilon - \mu^2 \left ( \delta + \eta \right )^2 + 
  i ( \delta + \eta ) \right ] \alpha_- - \left ( 2 |\alpha_{+}|^2 + 3 |\alpha_{-}|^2 \right ) \alpha_{-} -
    \alpha_{+}^2  \bar \alpha_{-}\, . \label{alphaMinus}
\end{eqnarray}
These are the expansions to third order of the
equations~(\ref{fpqeqn}--\ref{Neqn}) reduced to the centre manifold,
and are the equations that we study for the remainder of this paper.

\subsection{Symmetries}

The system~(\ref{alphaPlus},~\ref{alphaMinus}) is unchanged by
sign-reversal of either $\alpha_+$ or $\alpha_-$, \emph{i.e.} it is
equivariant with respect to the symmetry group $D_2 = Z_2(\sigma_x)
\times Z_2(\sigma_y)$, where we use the notation $Z_r(\xi)$ to
denote the cyclic group of order $r$ generated by the element
$\xi$. The actions of the group generators are:
\begin{equation}
 \sigma_x \, : \, 
  \left ( \begin{array}{c} \alpha_+ \\ \alpha_{-} \end{array} \right )
  \mapsto 
  \left ( \begin{array}{c} \alpha_+ \\ -\alpha_{-} \end{array} \right ) , 
   \hspace{20mm}
 \sigma_y \, : \, 
  \left ( \begin{array}{c} \alpha_+ \\ \alpha_{-} \end{array} \right )
  \mapsto 
  \left ( \begin{array}{c} -\alpha_+ \\ \alpha_{-} \end{array} \right ) ,
\end{equation}
and they represent the geometrical symmetries of the laser.
Using~(\ref{sincosmodes}) we see that, for example, $\sigma_x$
corresponds to a reflection with respect to the $x$-axis: this leaves
the cosine mode (amplitude $\alpha_+$) unaltered and changes the sign
of the sine mode (amplitude $\alpha_-$).  

The system~(\ref{alphaPlus},~\ref{alphaMinus}) is also equivariant
with respect to the $S^1$ action corresponding to global change of the
phases of the two amplitudes:
\begin{equation}
R_\varphi:
\left( \begin{array}{c}  \alpha_+ \\ \alpha_- \end{array} \right)
\mapsto
\left( \begin{array}{c}  e^{i\varphi}\alpha_+ \\ e^{i\varphi}\alpha_- 
                                \end{array} \right) .
\end{equation}
In other words, the amplitudes $\alpha_{+}$ and $\alpha_{-}$ are defined
only up to a common phase factor.  Therefore the total symmetry of
equations~(\ref{alphaPlus},~\ref{alphaMinus}) is the combination of
the geometrical and the phase invariance symmetry.  However, we note
that we can write
\begin{equation}
 \sigma_x = \sigma_y\circ R_\pi,
 \label{sx_From_sy}
\end{equation}
so the total symmetry group can be taken to be $Z_2 \times S^1$,
where $Z_2$ is generated by one of either $\sigma_x$ or $\sigma_y$.
The other geometrical symmetry can be obtained
using~(\ref{sx_From_sy}).  Therefore, the geometrical symmetry group
$D_2$ is recovered as a subgroup of $Z_2 \times S^1$.

We are now in a position to describe the bifurcation structure and
interaction of the two modes following the extensive analysis
in~\cite{dangelmayr91a}.  We summarise here the main results as they
apply to our case.  In the notation of ~\cite{dangelmayr91a} we have
$a=b=-1$ which places us on the lower boundary of region III(c) in
Figure 4 of ~\cite{dangelmayr91a}, from which the theoretical
bifurcation diagrams can then be deduced using Figures 6 and 7 of
~\cite{dangelmayr91a}.  In our Figure~\ref{fig:BifDiag} we give
versions of these diagrams as observed numerically in our context.
The next subsection is devoted to explaining this Figure.

\subsection{The case $\delta \ne 0$.}

We consider first the case $\delta >0$; the results obtained are also
valid for $\delta < 0$ after exchanging $\alpha_{+}$ with $\bar
\alpha_{-}$.  The important special case $\delta = 0$ of cavity
resonance, excluded from consideration in~\cite{dangelmayr91a}, will
be discussed in the next subsection.

The phase invariance of~(\ref{alphaPlus})~and~(\ref{alphaMinus})
implied by the $S^1$ symmetry makes the bifurcation analysis rather
cumbersome.  Therefore, it is convenient as in~\cite{dangelmayr91a} to
introduce the variables $A$, $\Phi$ and $\Theta$ defined by:
\begin{eqnarray}
 & v & = A \cos(\Phi/2) e^{i \Theta_1} , \\
 & w & = A \sin(\Phi/2) e^{i \Theta_2} , \\
 & \Theta & = \Theta_1 + \Theta_2 , 
\end{eqnarray}
where
\begin{equation}
 v = \alpha_+ - i \alpha_- 
 \quad \mbox{and} \quad
 w = \bar \alpha_+ - i \bar \alpha_- .
 \label{def_vw}
\end{equation}
We can reduce the two complex equations~(\ref{alphaPlus})
and~(\ref{alphaMinus}) to three real equations for $A$, $\Phi$ and
$\Theta\,$:
\begin{eqnarray}
 & \displaystyle \frac{d A}{d t} = &
  \left [ 
   \varepsilon - \mu^2 \left ( \delta^2 + \eta^2 \right ) 
  \right ] A - \left [ 1 + \frac{1}{2} \sin^2\Phi \right ] A^3 + 
   c A \sin\Phi \cos\Theta \cos\alpha \nonumber \\
 & & \qquad \label{A_eq} \\
 & \displaystyle \frac{d \Phi}{d t} = &-\frac{A^2}{2} \sin(2 \Phi) +
  2 c \left [ 
  \cos\Theta \cos\alpha \cos\Phi - \sin\Theta \sin\alpha 
  \right ] \label{Phi_eq} \\
 & \displaystyle \frac{d \Theta}{d t} = & 
   - 2 \frac{c}{\sin\Phi}
  \left [ 
   \sin\Theta \cos\alpha + \cos\Theta \sin\alpha \cos\Phi
  \right ] \label{Theta_eq}
\end{eqnarray}
({\it cf.} \cite{dangelmayr91a}) where $c$ is defined by
\begin{equation}
 c e^{i \alpha} = \eta ( 2 \mu^2 \delta - i ) 
 \label{epsDK}
\end{equation}
and plays the r\^ole of a (complex) symmetry breaking parameter in
that it is zero if and only if the $O(2)$ symmetry of the system is
present.  Note that in terms of $A$, $\Phi$ and $\Theta$ both
$\sigma_x$ and $\sigma_y$ have the same realisation
\begin{equation}
 \sigma_x = \sigma_y = \sigma :
 \left ( \begin{array}{c}
  A \\ \Phi \\ \Theta 
 \end{array} \right ) \mapsto
 \left ( \begin{array}{c}
  A \\ \pi - \Phi \\ -\Theta 
 \end{array} \right )
 \label{sigma_symm}
\end{equation}
thus highlighting the fact that the symmetry of the problem can
ultimately be considered to be $Z_2$, once the phase invariance
$S^1$ symmetry is factored out.  
 
\begin{figure}
 \centerline{\epsfig{file={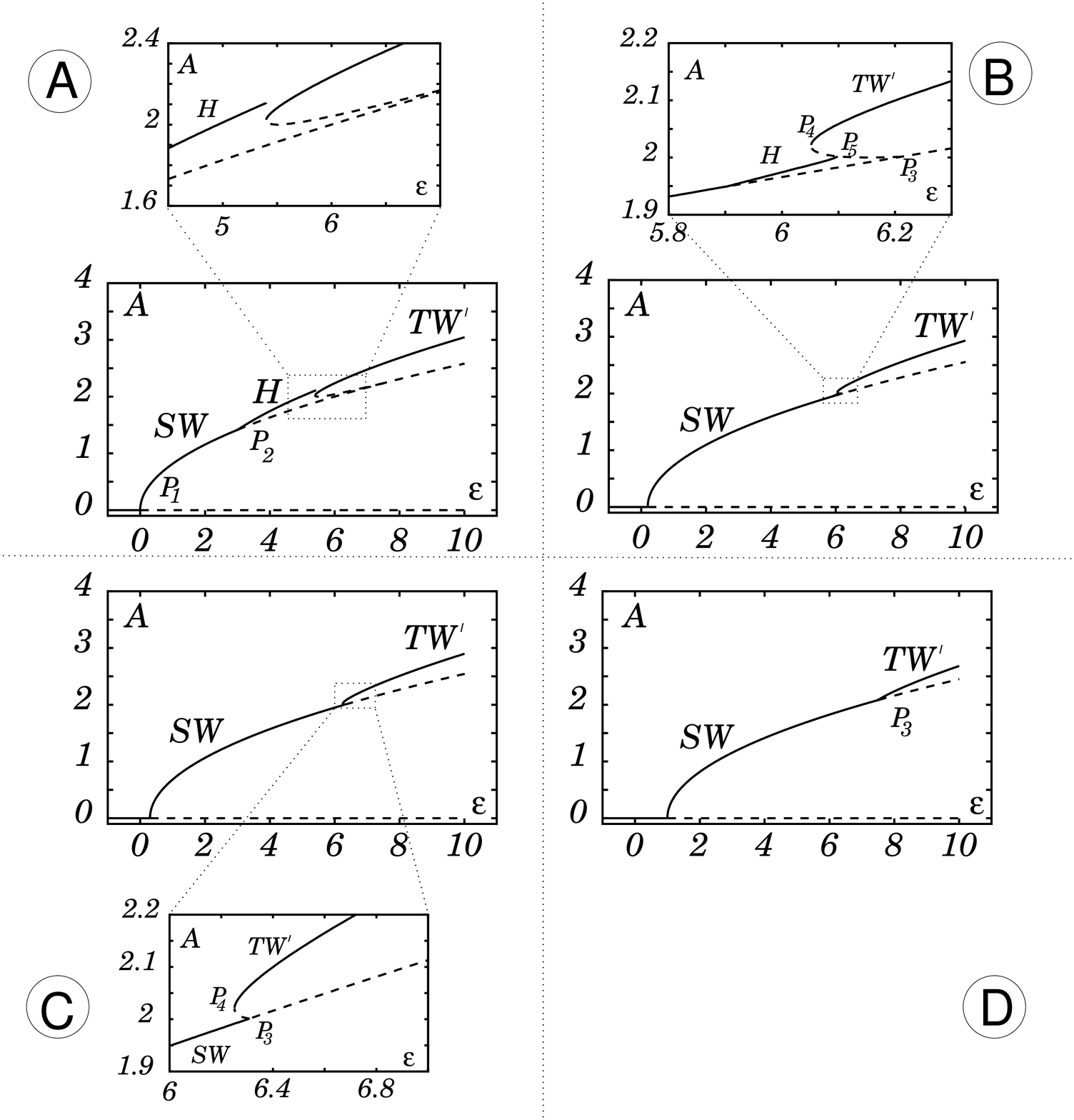},width=130mm}}
 \caption{\label{fig:BifDiag} \it Bifurcation diagrams for $\kappa = 1$
 and $\eta=1$ (which imply that $\delta_2=2$) and different values of
 the detuning parameter: $\delta = \{1, 1.9, 2.1, 3\}$ from (A) to
 (D).  The solid (dashed) lines correspond to stable (unstable)
 solutions.  We have not drawn the unstable branch $SW_\pi$.  The
 bifurcation points $P_j$, $j=1,..,5$ have horizontal coordinate
 $\varepsilon_j$.  The vertical coordinate of the branches of
 stationary solutions is their amplitude $A$.  That of the Hopf branch
 is the maximum of the amplitude during one period.  For this reason
 in panel (A) the Hopf branch does not touch the limit point.}
\end{figure}

At $\varepsilon = \varepsilon_1 \equiv \mu^2 (\delta - \eta)^2$ the
trivial solution $\alpha_{\pm} = 0$ ($A=\Theta=0$, $\Phi=\pi/2$) loses
its stability through a Hopf bifurcation (bifurcation point $P_1$ in
Figure~\ref{fig:BifDiag}).  A new nonzero single mode solution
(standing wave solution $SW$ in the notation of~\cite{dangelmayr91a}
and in Figure~\ref{fig:BifDiag}) with frequency
\begin{equation}
 \omega_{+} = \delta - \eta,
\end{equation}
and amplitudes
\begin{equation}
 |\alpha_{+}|^2 = \frac{1}{3} \left ( \varepsilon - 
    \mu^2 \omega_+^2 \right ) , 
 \hspace{20mm} \alpha_{-} = 0 
 \label{SolSW}
\end{equation}
appears and is stable.  Note that in the context of
equations~(\ref{A_eq}--\ref{Theta_eq}) this bifurcation is a pitchfork
and the new stable solution is stationary with values
\begin{equation}
 A^2 = \frac{2}{3} \left ( \varepsilon - \mu^2 \omega_+^2 \right ), 
 \quad \Phi = \pi/2, \quad \Theta=0.
\end{equation}
This simplification is a result of having factored out the $S^1$ (phase
invariance) group.   The stationary point is a focus for 
\begin{equation}
 \varepsilon < \varepsilon_{N} \equiv 6 \eta + \mu^2 (\delta-\eta)^2 
\end{equation}
and is a node otherwise.

There is an additional bifurcation from the zero solution.  At
$\varepsilon = \varepsilon_6 \equiv \mu^2 (\delta+\eta)^2$ there is a
Hopf bifurcation to the unstable solution ($SW_\pi$ in the
notation of~{\cite{dangelmayr91a})
\begin{equation}
 \alpha_+ = 0, \quad 
 |\alpha_-|^2 = \frac{1}{3} 
  \left [ \varepsilon - \mu^2(\delta+\eta)^2 \right ] .
 \label{SolSWpi}
\end{equation}
In terms of $A$, $\Phi$ and $\Theta$ this is a pitchfork bifurcation
to the unstable solution
\begin{equation}
 A^2 = \frac{2}{3} \left [ \varepsilon - \mu^2(\delta+\eta)^2 \right ] ,
 \quad
 \Phi = \pi/2 , \quad \Theta = \pi . 
\end{equation}
There are no further bifurcations on this branch.

The $SW$ branch may have a secondary Hopf bifurcation to
a two-frequency solution (H in Figure~\ref{fig:BifDiag}) as the
bifurcation parameter is increased to the value
\begin{equation}
 \varepsilon_2 \equiv \mu^2 
  \left [ ( \delta + \eta )^2 + 8 \delta \eta \right ] .
\end{equation}
This solution is characterised by both modes having non-zero
amplitude, but different frequency so that beating between the two
modes occurs.  

As $\delta$ tends to zero $\varepsilon_2$ tends to $\varepsilon_1$, so
that the secondary Hopf instability occurs closer and closer to the
primary bifurcation from the trivial solution (see panel (A) in
Figure~\ref{fig:BifDiag}).  This secondary Hopf bifurcation point
$P_2$ exists only in the range
\begin{equation}
 0 < \delta < \delta_2 \equiv \frac{1}{2 \mu^2}\,.
\end{equation}
In fact, another two stationary solutions appear at a third
bifurcation point $P_3$ on the $SW$ branch, at 
\begin{equation}
 \varepsilon = \varepsilon_3 \equiv \frac{3 \eta}{2 \mu^2 \delta} +
  \mu^2 \left [ \eta^2 + \delta^2 + 4 \delta \eta \right ] \,.
\end{equation}
In these solutions the two modes are frequency locked to each other
and thus the intensity pattern is stationary.

If $\delta = \delta_2$ then $\varepsilon_2 = \varepsilon_3$ and the
two bifurcation points $P_2$ and $P_3$ collide.  The solutions that
appear at $P_3$ ($TW'$ in Figure~\ref{fig:BifDiag}) have $\alpha_{+}$
and $\alpha_{-}$ both nonzero ($\Phi \ne \pi/2$ and $\Theta \ne 0$)
and are interchanged by the symmetry operation $\sigma$ defined
by~(\ref{sigma_symm}).  The bifurcation is sub-critical if
\begin{equation}
 0 \le \delta \le \delta_4 \equiv 
   \sqrt{\frac{3}{2}} \delta_2 , 
\end{equation}
(see panels (B) and (C) of Figure~\ref{fig:BifDiag}) and has a limit
point $P_4$ at
\begin{equation}
 \varepsilon \equiv \varepsilon_4 = 2 \sqrt{6} \eta +
  \mu^2 (\eta^2 + \delta^2) .
\end{equation}
If $\delta = \delta_4$ then $\varepsilon_3 = \varepsilon_4$ and the
limit point $P_4$ hits the $SW$ branch at $P_3$.  For larger values of
$\delta$ the bifurcation is supercritical (see panel (D) of
Figure~\ref{fig:BifDiag}).

\begin{figure}
 \centerline{\epsfig{file={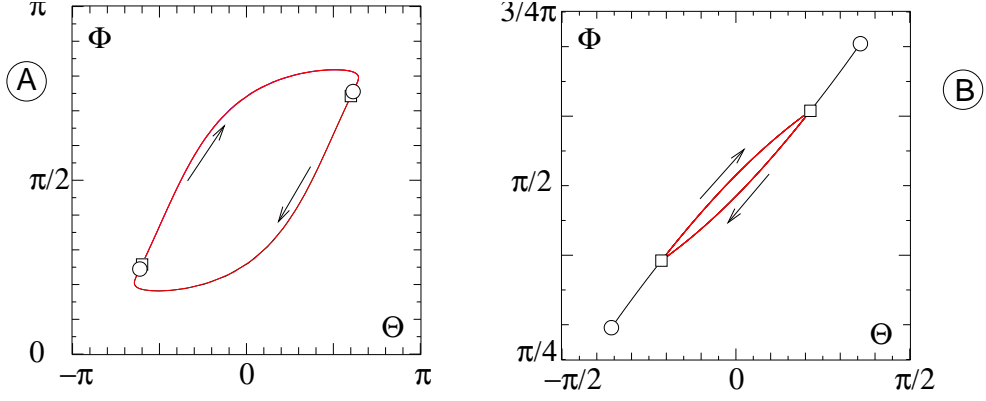},width=130mm}}
 \caption{\label{fig:BlueSky} \it (A) Blue sky bifurcation that ends
 the existence of the periodic orbit in panel (A) of
 Figure~\ref{fig:BifDiag} at $\varepsilon=5.399$.  Two saddle node
 bifurcations (circle: stable fixed point, square: unstable fixed
 point) appear at symmetrical locations along the periodic orbit.  (B)
 Disappearance of the periodic orbit in the case of panel (B) of
 Figure~\ref{fig:BifDiag} at $\varepsilon=6.098$.  The periodic orbit
 clashes against the heteroclinic connections between the two unstable
 fixed points.  All the parameters are as in
 Figure~\ref{fig:BifDiag}.}
\end{figure}

The secondary Hopf branch ends either at $P_4$ or, for larger values
of $\delta$, at a point $P_5$ on the the unstable branch that joins
$P_3$ and $P_4$ (see panels (B) and (C) of Figure~\ref{fig:BifDiag}).
In the former case the periodic orbit disappears in a blue sky
bifurcation (panel (A) of Figure~\ref{fig:BlueSky}), that is the
period tends to infinity until a saddle-node bifurcation creates a
sink and a saddle that `block' the periodic orbit:
compare~\cite[Fig.3]{wieczorek99a}.  In the latter case the period
also tends to infinity, but the orbit collides with the heteroclinic
orbits that join the unstable fixed points in the $TW'$ branch (see
panel (B) of Figure~\ref{fig:BlueSky}).  There is no analytical
expression for the coordinates of $P_5$ and we have used
AUTO~\cite{doedel91a,doedel91b} to plot it in
Figure~\ref{fig:BifDiag}.  Finally, note that even though the point
$P_3$ tends to infinity as $\delta$ tends to zero, $P_4$ is always
finite: therefore the periodic orbit exists for only a finite range of
values of $\varepsilon$.  These results are summarised in the state
diagram in Figure~\ref{fig:StateDiag}.

\begin{figure}
 \centerline{\epsfig{file={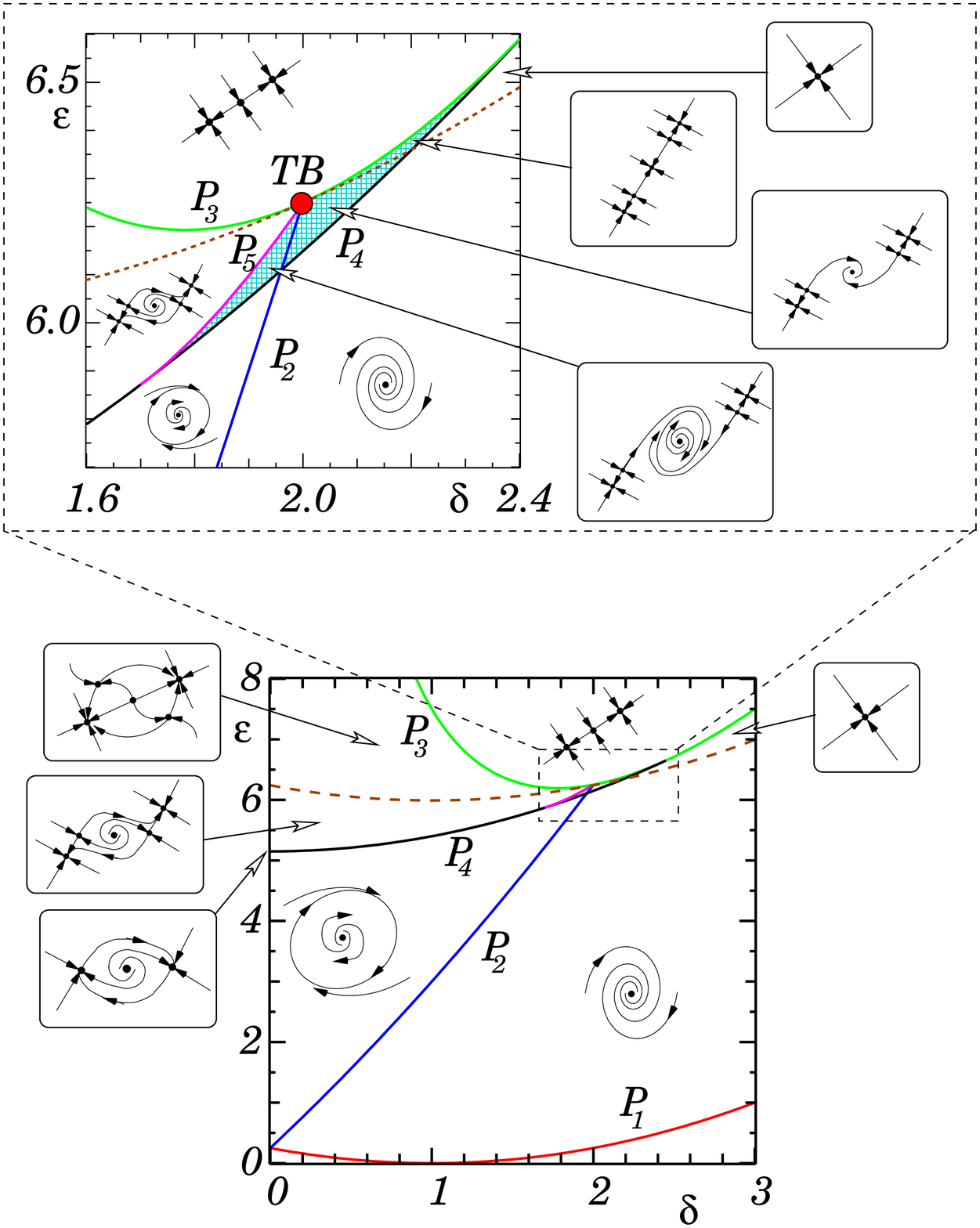},width=130mm}}
 \caption{\label{fig:StateDiag} \it State diagram of the
 equations~(\ref{A_eq}--\ref{Theta_eq}) for $\kappa=1$ and
 $\eta=1$.  The line marked $P_j$ represents the bifurcation point
 $P_j$ of Figure~\ref{fig:BifDiag} as a function of the parameters
 $\varepsilon$ and $\delta$.  The dashed line represents the curve
 $\varepsilon_{N}$: below it the $SW$ point is a focus, above it is a
 node.  The large dot next to the letters TB indicates the
 Takens-Bogdanov bifurcation point.  The top part of the figure is an
 enlargement of the rectangular dashed section in the bottom part.
 The orbits drawn are just indicative of the type of dynamics expected
 in the corresponding parameter region and are not intended to be in
 any particular plane of the $A$, $\Phi$ and $\Theta$ space.}
\end{figure}

The state diagram is organised around the bifurcation point of highest
codimension (two), namely the $Z_2$-symmetric Takens-Bogdanov point
(indicated by TB in Figure~\ref{fig:StateDiag}) which occurs when
the points $P_2$, $P_3$ and $P_5$ (see panels A and B of
Figure~\ref{fig:BifDiag}) coincide, \emph{i.e.} at $\delta = \delta_2$.
In~\cite{dangelmayr91a} the normal forms in the neighbourhood of this
bifurcation point are derived as
\begin{eqnarray}
 & & \dot \xi = \eta + O(5) , \label{TB_eq1} \\
 & & \dot \eta = M_{TB} \xi^3 + N_{TB} \xi^2 \eta + O(5) ,
      \label{TB_eq2} 
\end{eqnarray}
in suitable coordinates $\xi$ and $\eta$.  The symbol $O(5)$
represents terms of order at least five in $\xi$ and $\eta$ and the
coefficients $M_{TB}$ and $N_{TB}$ depend on the parameters of
equations~(\ref{alphaPlus},~\ref{alphaMinus}).  The condition for a
(non degenerate) Takens-Bogdanov singularity is $M_{TB} N_{TB} \ne 0$.
Using the calculations given in Appendix A of \cite{dangelmayr91a} we
find in our case $M_{TB}=1/3$ and $N_{TB}=\mp 25/(18 \sqrt{2})$, where
the signs refer respectively to the cases $\delta > 0$ and $\delta <
0$.  This implies that the bifurcation geometry in a neighbourhood of
the Takens-Bogdanov point in Figure~\ref{fig:StateDiag} is
topologically equivalent to that obtained from a versal unfolding of a
standard model as described for example in~\cite{takens74a}.

The bifurcation geometry in the neighbourhood of a $Z_2$-symmetric
Takens-Bogdanov point is well documented: initial studies by
Takens~\cite{broer01a} with further analysis given by
Carr~\cite{carr81a}, Guckenheimer and Holmes~\cite{guckenheimer1983},
Arnol'd~\cite{arnold83a}, Khorozov~\cite{khorozov85a} (see
also~\cite{arnold94a}) and others show that a versal unfolding of the
vector field at this singularity gives a bifurcation diagram
determined by a smooth curve passing through the Takens-Bogdanov point
and two rays emanating from it not tangent to the curve or each other
(in contrast to the non-symmetric case).  This is indeed the
description of the bifurcation diagram in the neighbourhood of the
point TB in Figure~\ref{fig:StateDiag}.

In applications, a Takens-Bogdanov bifurcation, whether symmetric or
not, is often accompanied by additional characteristic features that
arise from interaction with further equilibrium states.  This can be
seen, for example, in the forced van der Pol
equation~\cite{gillies54a,holmes78a} (see
also~\cite{guckenheimer1983}) and in the dynamics of bulk liquid
crystals in a shear flow~\cite{vicentealonso01a}. The explanation for
these accompanying patterns lies in the geometry of unfolding of a
singularity of higher degeneracy and thus higher codimension: although
it does not appear among the vector fields studied, its presence
behind the scenes {\it organises} the configuration of lower codimension
bifurcations.  In the case of the van der Pol oscillator there is a
triangular region in the neighbourhood of the (non symmetric)
Takens-Bogdanov point (see Figure 2.1.2 of~\cite{guckenheimer1983})
whose footprint can recognised as part of the codimension-3 non
symmetric bifurcations studied by Dumortier {\it et al.}~\cite{dumortier91a}.

An analogous phenomenon occurs here.  The organising centre is a
degenerate $Z_2$-symmetric Takens-Bogdanov singularity involving
coalescence of \emph{five} equilibrium states (instead of three for
the non-symmetric case) and arising when $M_{TB}=0$ in
equation~(\ref{TB_eq2}).  As in the van der Pol case, there is a
`triangular' region (shaded area in Figure~\ref{fig:StateDiag}) whose
features are determined by the interactions of the five fixed points.
However, we are not aware of a rigorous treatment of this case even
though many authors have touched on aspects of it.  For example,
interactions of the five equilibria lead to interesting dynamical
phenomena in certain fluid dynamical convection problems
\cite{dacosta81a}.  In the case of lasers, L\'opez-Ruiz \emph{et
al.}~\cite{lopezruiz94a} give a bifurcation analysis only in the
neighbourhood of the symmetric Takens-Bogdanov point.  Krauskopf
\emph{et al.}~\cite{wieczorek99a,wieczorek00a,krauskopf00a,degen00a}
give spectacular and informative bifurcation diagrams away from the
Takens-Bogdanov point, for a model that does not, however, involve the
$Z_2$-symmetry.  Rucklidge~\cite{rucklidge01a} considers bifurcations
only near the Takens-Bogdanov point and with the sign of $M_{TB}$
opposite to ours.

\subsection{The case $\delta=0$}

\subsubsection{Introduction}

\begin{figure}
 \centerline{\epsfig{file={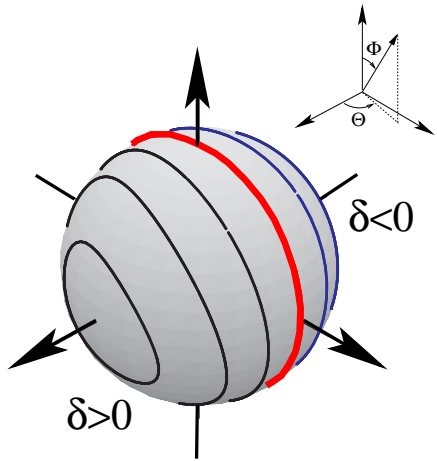},width=60mm}}
 \caption{\label{fig:delta=0} \it Representation of the periodic
 orbits on the equations~(\ref{A_eq}--\ref{Theta_eq}) in
 ($\Theta,\Phi$)-spherical coordinates for different values of the
 detuning parameter $\delta$.  The orbits are contained in the nearer
 (further) hemisphere if $\delta>0$ ($\delta<0$).  The limiting case
 $\delta=0$ corresponds to an orbit coinciding with a meridian of the
 sphere (thick line).  In these simulations $\kappa=1$, $\eta=1$ and
 $\delta=\{0.75, 0.50, 0.25, 0.00, -0.25, -0.50, -0.75\}$.}
\end{figure}

If $\delta=0$ the symmetry-breaking parameter $c$
defined by equation~(\ref{epsDK}) is purely imaginary and equal to $-i
\eta$.  The physical interpretation is that from the point of view of
the laser the modes are distinct because they have different resonant
frequencies ($\eta \ne 0$), but also there is no reason for the laser
to chose one rather than the other as the pump parameter is increased
through threshold.  If instead $\delta \ne 0$ then one mode has a
lower threshold than the other, reflected in the fact that the
symmetry-breaking parameter has a non-zero real part.

The solutions of equations~(\ref{A_eq}-\ref{Theta_eq}) can be
conveniently represented in spherical coordinates: the variable
$\Theta$ plays the r\^ole of the polar angle and the variable $\Phi$
that of the azimuthal angle (see Figure~\ref{fig:delta=0}).  The two
fixed points $A \ne 0$, $\Phi=\pi/2$ and $\Theta=\{0,\pi\}$ correspond
to diametrically opposed points in the equatorial plane and the
periodic orbits are closed loops around them.  Moreover, the sign of
the detuning $\delta$ determines in which hemisphere the periodic
orbit is situated.  The orbits for opposite values of $\delta$
are related by the symmetry $\sigma$ defined by~(\ref{sigma_symm}).

If $\delta=0$ the orbit is constrained to the meridian $\Theta=\pi/2$
that separates the two hemispheres (see Figure~\ref{fig:delta=0}) and
thus has higher symmetry.  This suggests in turn that the model may
have an additional symmetry if $\delta=0$.  Therefore if we identify
this additional symmetry we can use the tools of symmetric bifurcation
theory \cite{golubitsky1988,chossat00a} to analyse the $\delta=0$
case and its response to symmetry-breaking perturbations.  This is the
approach we follow throughout the rest of this section.

\subsubsection{Additional discrete symmetry}

As a first step we identify the symmetry group of the
equations~(\ref{alphaPlus},~\ref{alphaMinus}) when $\delta=0$.  In this
case the equations~(\ref{alphaPlus},~\ref{alphaMinus}) become
\begin{eqnarray}
 \begin{array}{l}
  \dot\ap=(\lambda-i\eta)\ap-(3|\ap|^2+2|\am|^2)\ap-\bar\ap\am^2\\
  \dot\am=(\lambda+i\eta)\am-(2|\ap|^2+3|\am|^2)\am-\bar\am\ap^2
 \end{array}
 \label{d=0}
\end{eqnarray}
where $\lambda=\eps-\mu^2\eta^2$.  Under the coordinate change~(\ref{def_vw})
these correspond to the perturbed equations (2.7) of \cite{dangelmayr91a}
\begin{equation}
 \begin{array}{l}
  \dot v=g(\lambda,|v|^2,|w|^2)v + c \bar w\\ 
  \dot w=\bar g(\lambda,|v|^2,|w|^2)w + \bar c \bar v
 \end{array}
 \label{dangeqs}
\end{equation}
which have $Z_2\times S^1$ symmetry given by 
\begin{equation}
 \left ( \begin{array}{c} v \\ w \end{array} \right ) 
  \mapsto 
  \left ( \begin{array}{c} \bar{w} \\ \bar{v} \end{array} \right ) 
  \quad \mbox{and} \quad
 \left ( \begin{array}{c} v \\ w \end{array} \right ) 
  \mapsto 
  \left ( \begin{array}{c} e^{i\varphi}v \\ 
           e^{-i\varphi}w \end{array} \right )
  \, ,
\end{equation}
as shown earlier using equation~(\ref{sx_From_sy}).  However, if
$\delta=0$ the perturbation coefficient $c$ is purely imaginary, and
this case is excluded from consideration in \cite{dangelmayr91a}.
Since $g$ is real there is now an additional symmetry of order 2:
\begin{equation}
 \tau: \left ( \begin{array}{c} v \\ w \end{array} \right ) 
  \mapsto -i 
       \left ( \begin{array}{c} \bar v \\ \bar w \end{array} \right )
\end{equation}
or, equivalently,
\begin{equation}
 \tau:\left ( \begin{array}{c} \ap \\ \am \end{array} \right ) \mapsto 
      \left ( \begin{array}{c} \bar\am \\ \bar\ap \end{array} \right )
      \, .
\end{equation}
Combined with the $S^1$ action 
\begin{equation}
 R_\varphi \,: \, 
  \left ( \begin{array}{c} \ap \\ \am \end{array} \right )
  \mapsto 
  e^{i\varphi} \left ( \begin{array}{c} \ap \\ \am \end{array} \right ).
\end{equation}
this generates an action of $O(2)$.   Moreover, the symmetry
\begin{equation}
 \rho : \left ( \begin{array}{c} \ap \\ \am \end{array} \right ) 
  \mapsto \left ( \begin{array}{c} i\ap \\ -i\am \end{array} \right )
\end{equation}
of order 4 commutes with both $\tau$ and $R_\varphi$.  This therefore
gives as total symmetry group the subgroup $Z_4 \times O(2)$ of the
group $S^1 \times O(2)$ generated by
\begin{equation}
 P_\theta : \left ( \begin{array}{c} \ap \\ \am \end{array} \right ) 
  \mapsto \left ( \begin{array}{c} e^{i\theta} \ap \\ 
                  e^{-i\theta} \am \end{array} \right )
\end{equation}
and $O(2)$ as above.  Notice that the $S^1\times O(2)$ action here
corresponds to that in~\cite{dangelmayr91a} under direct substitution
of $v,w$ in place of $\alpha_+,\alpha_-$ although in our problem these
coordinates are in fact related by~(\ref{def_vw}).  We shall return to
this significant point later.

Observe that there is redundancy in the group action since the element
$\zeta=\rho^2 \, R_\pi$ acts trivially.  Removing this redundancy
yields the quotient group $(Z_4\times O(2))/Z_2(\zeta)$ which is
cumbersome to work with both conceptually and notationally -- in
contrast to the $(D_2\times S^1)/Z_2(\sigma_x)$ case handled earlier
which reduced simply to $Z_2\times S^1\,$.  We therefore prefer to
focus on the group $Z_4\times O(2)$ with its transparent algebraic
structure, while recognising that its \emph{action} on ${\mathbb C}^2
\cong {\mathbb R}^4$ is not faithful.
 
\begin{figure}
 \centerline{\epsfig{file={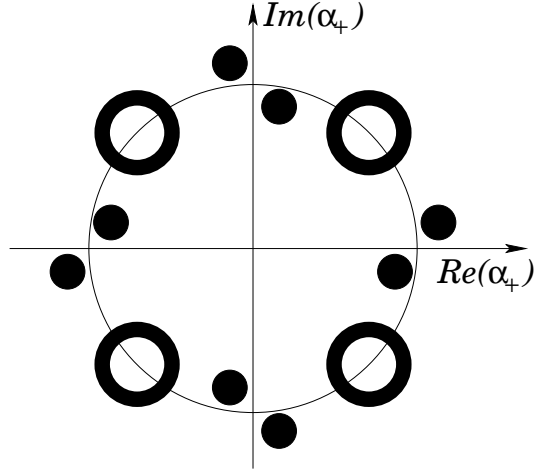},width=70mm}}
 \caption{\label{fig:FixedSpaces} \it Schematic representation of
 the intensity patterns in $\fix(\Sigma_1)$,
 equation~(\ref{intensity_FixS1}).   The black areas represent regions
 of high intensity.   Their position along the thin circle indicates the
 corresponding value of the argument of $\alpha_+$.}
\end{figure}

Once the symmetry group of the dynamical system has been established,
we proceed to identify its subgroups and their fixed point spaces.
The fixed point space $\fix(\Sigma)$ of a given subgroup $\Sigma$ is
the vector space whose points are left unchanged by the action of
$\Sigma\,$.  The importance of fixed point spaces for bifurcation
analysis is that they are invariant under the dynamics.  In particular
if $\fix(\Sigma)$ is one-dimensional then typically at bifurcation
there must be a branch of stationary solutions lying in $\fix(\Sigma)$
(Equivariant Branching Lemma) and thus exhibiting the symmetries
$\Sigma\,$.  If $\fix(\Sigma)$ is two-dimensional and the eigenvalues
at bifurcation are purely imaginary then typically there is a branch
of periodic solutions lying in that subspace (Equivariant Hopf
Theorem).  See~\cite{golubitsky1988} or~\cite{chossat00a} for precise
statements of these results.

The subgroups of $Z_4\times O(2)$ with 2-dimensional fixed
point subspaces are found among those for $S^1\times O(2)$
(see~\cite{golubitsky1988}) and are (up to conjugacy) as follows:
\begin{equation}
 \Sigma_1 \, : \, Z_2(\tau) \times Z_2(\rho^2 \, R_\pi) , \qquad
 \Sigma_2 \, : \, Z_4(\rho \, R_{-\frac{\pi}{2}}) .
\end{equation}
Their respective fixed point spaces are
\begin{equation}
 \fix(\Sigma_1) = \left \{ \left ( 
   \begin{array}{c} \ap \\ \bar\ap \end{array} \right ) \right \} , \qquad
 \fix(\Sigma_2) = \left \{ \left (
   \begin{array}{c} \ap \\ 0 \end{array} \right ) \right \} \, .
\end{equation}
We next identify these fixed point spaces in terms of laser patterns.
We know from equation~(\ref{sincosmodes}) that $\alpha_+$ and
$\alpha_-$ are the amplitudes of a cosine and sine mode respectively.
In particular, for the purpose of this example we can assume that
$m=1$ in (\ref{sincosmodes}) so that the expansion of the electric
field can be written as
\begin{equation}
 F(r,\psi,t) = \alpha_+(t) A_0(r) \cos(\psi) + 
               \alpha_-(t) A_0(r) \sin(\psi) .
\end{equation}
The intensity pattern that corresponds to the elements of
$\fix(\Sigma_1)$ is
\begin{equation}
 | F_{\fix(\Sigma_1)} |^2 = |A_0(r)|^2 |\alpha_+|^2 
   \left \{ 1 + \cos \left [ 2 \mbox{arg}(\alpha_+) \right ]
            \sin(2 \psi) \right \} .
 \label{intensity_FixS1}
\end{equation}
As the argument of $\alpha_+$ goes from $0$ to $2 \pi$ the intensity
pattern changes according to Figure~\ref{fig:FixedSpaces}.  The
elements of $\fix(\Sigma_2)$ have an intensity pattern given by
\begin{equation}
 | F_{\fix(\Sigma_2)} |^2 = |A_0(r)|^2 |\alpha_+|^2 \cos^2(\psi) .
\end{equation}
This corresponds to two bright spots aligned along the horizontal axis,
independently of the phase of $\alpha_+$.  There is also the
corresponding sine pattern: this corresponds to $\fix(\Sigma_2')$
where $\Sigma_2'$ is the conjugate $\Sigma_2'= \tau\Sigma_2 \tau^{-1}$
of $\Sigma_2\,$.
\medskip

Given a stationary or periodic solution of a symmetric dynamical
system we obtain other solutions by applying the elements of the
symmetry group to it.  It is important therefore, to understand how
the full group $Z_4\times O(2)$ acts on the fixed point spaces
$\Sigma_1$ and $\Sigma_2$.

Let $\ab$ denote the vector $(\alpha_+, \alpha_-)^T$.  If
$0\ne\ab\in\fix(\Sigma_1)$ then $\rho\ab\in\fix(\Sigma_1)$ as $\rho$
commutes with $\tau$.  On the other hand, $R_\varphi \ab \in
\fix(\Sigma_1)$ just when $\varphi=0,\pi$.  Therefore the $Z_4\times
O(2)$ orbit of $\ab$ consists of two circles intersecting
$\fix(\Sigma_1)$ at $\{\ab,-\ab\}$ and $\{\rho\ab,-\rho\ab\}$.  In
physical terms this result means that if there is a solution in
$\fix(\Sigma_1)$, then there are infinitely many copies of it, each
obtained by multiplication by a phase factor.  However, if the phase
factor is equal to $\pi$ then the net effect is not to obtain a
different solution but to have moved half a period in
$\fix(\Sigma_1)$.

If $0\ne\ab\in\fix(\Sigma_2)$ then $\rho\ab\in\fix(\Sigma_2)$ and
$R_\varphi\ab\in\fix(\Sigma_2)$ so the $Z_4\times O(2)$ orbit of $\ab$
consists of two circles $C$ and $\tau C$ lying in $\fix(\Sigma_2)$ and
$\fix(\Sigma_2')$ respectively.  In other words, the net effect of the
group action is to multiply a sine or cosine solution by a phase
factor.

We can combine these results and make use of the Equivariant Hopf
Theorem \cite{golubitsky1988,chossat00a} to deduce the existence of
branches of periodic solutions with these symmetries that bifurcate
from the trivial solution.

\begin{theorem}\label{bifs}
For the equations~(\ref{d=0}) there exist (at least) the following
periodic solutions branching from the trivial solution as $\lambda\/$
passes through zero:

\begin{enumerate}

\item a family of solutions of the form
\begin{equation}
 W_{\varphi}(t) = 
  \left ( \begin{array}{c} 
   r_\lambda e^{i\varphi} e^{i(\eta t+\theta)} \\*[3mm]
   r_\lambda e^{i\varphi} e^{-i(\eta t+\theta)}  
  \end{array} \right ) ,
 \label{W_phi}
\end{equation}
where $r_\lambda\sim\lambda^\frac{1}{2}$ and $\varphi$ and $\theta$
are arbitrary.  Here $W_{\varphi}$ has reflection symmetry under
$R_\varphi \tau R_\varphi^{-1} = R_{2 \varphi} \tau$ and a $Z_4$
spatio-temporal symmetry under which increasing time by $\pi/(2\eta)$
is equivalent to applying $\rho\,$;

\item a pair of solutions of the form 
\begin{equation}
 V_+\,,V_- = 
  \left ( \begin{array}{c} 
   r_\lambda e^{i(\eta t+\theta)} \\ 0 
  \end{array} \right )\, , 
  \left ( \begin{array}{c} 
   0 \\ r_\lambda e^{-i(\eta t+\theta)} 
  \end{array} \right )
\end{equation}
respectively, where $r_\lambda \sim\lambda^\frac{1}{2}$ and $\theta$
is arbitrary.  Each solution has spatio-temporal symmetry
$\widetilde{SO}(2)$ given by shifting time by $\Delta t$ and rotating
by $\theta=-\eta \Delta t$.

\end{enumerate}

\end{theorem}

\noindent
Under time evolution the solution $W_{\varphi}$ cycles across all the
patterns in Figure~\ref{fig:FixedSpaces}.  The solutions $V_{+}$ and
$V_{-}$ correspond to the solutions $SW$ and $SW_\pi$,
equations~(\ref{SolSW}) and~(\ref{SolSWpi}).  We will see below that
for our particular system~(\ref{d=0}) these are the only solutions
that branch from the trivial solution and, moreover, that the
$W_{\varphi}$ solution is stable and the $V_\pm$ are unstable.

\subsubsection{Disguised rotational symmetry}

Interestingly as it turns out, the relations between the solutions for
$\delta=0$ and those for $\delta \ne 0$ described above can be
elucidated by rewriting equations~(\ref{d=0}) in terms of new and
somewhat artificial variables that restore to the equations a higher
degree of symmetry.  The linear terms of~(\ref{d=0}) commute with
rotational symmetries
\begin{equation}
 \left ( \begin{array}{c} \ap \\*[1mm] \am \end{array} \right ) 
  \mapsto 
  \left ( \begin{array}{c} 
   e^{i\varphi_1} \ap \\*[1mm] e^{i\varphi_2}\am 
  \end{array} \right )
 \label{indepPhasesSym}
\end{equation}
for arbitrary $\varphi_1,\varphi_2$, and so Takens' theory of normal
forms~\cite{takens74a} suggests the possibility of using a nonlinear
coordinate change to remove (to any desired order) all nonlinear terms
that do not share this symmetry.  It is important to stress that the
change of variables is nonlinear: although the linear
transformation~(\ref{def_vw}) removes the non-$O(2)$-symmetric cubic
terms $\bar \alpha_+ \alpha_-^2$ and $\alpha_+^2 \bar \alpha_-$ from
equations~(\ref{d=0}), the linear terms in $v$ and $w$ in the
resulting equations~(\ref{dangeqs}) at the same time lose their full
rotational symmetry.  Normal form theory states that the nonlinear
transformation will succeed when there are no {\it resonances} (that
is, linear relations with integer coefficients) between the
eigenvalues at the bifurcation point. In our case the eigenvalues at
the bifurcation point are $\pm i \eta$ and there are of course strong
resonances, but careful inspection of the normal form method shows
that the terms $\bar\ap\am^2$ and $\bar\am\ap^2$ can nevertheless be
removed.  To be precise:

\begin{theorem}
The nonlinear coordinate transformation
\begin{equation}
 \left ( \begin{array}{c} \ap \\*[1mm] \am \end{array} \right ) = 
  \left ( \begin{array}{c} \bp \\*[1mm] \bm \end{array} \right ) - 
   \frac{1}{2 \lambda + 4 i \eta}
   \left ( \begin{array}{c} \bar\bp\bm^2 \\*[1mm]
     \bar\bm\bp^2 \end{array} \right )
 \label{betaTransf}
\end{equation}
converts the equations~(\ref{d=0}) into
\begin{equation}
 \begin{array}{l}
  \dot\bp=(\lambda-i\eta)\bp-(3|\bp|^2+2|\bm|^2)\bp \\
  \dot\bm=(\lambda+i\eta)\bm-(2|\bp|^2+3|\bm|^2)\bm 
 \end{array}
 \label{betaEq}
\end{equation}
up to order 3.
\end{theorem}

\emph{Proof}.\enspace Direct verification, or appeal to general
theory~\cite{takens74a} and the observation that it is the difference
and not the sum of the eigenvalues $\lambda\pm i\eta$ whose vanishing
provides the obstruction to removing the non-rotational cubic terms.

Observe that the equations~(\ref{betaEq}) have the
form~(\ref{dangeqs}) with $e=0$ and with $v,w$ corresponding to
$\beta_+,\beta_-\,$, in contrast to the original correspondence
defined by~(\ref{betaTransf}) and~(\ref{def_vw}).  Using the framework
of~\cite{dangelmayr91a} we can therefore express the symmetry as
follows:

\begin{corollary}
After changing to  $\bb \equiv (\beta_+, \beta_-)^T$  coordinates and
truncating at order 3 the system~(\ref{d=0}) has $S^1\times O(2)$
symmetry with $O(2)$ generated by $\tau,R_\varphi$ and with
$S^1$-action $P_\theta$.
\end{corollary}

\emph{Remark 1}.\enspace We recover the independent phase
symmetries~(\ref{indepPhasesSym}) by taking $\varphi_1=\varphi+\theta$
and $\varphi_2=\varphi-\theta\,$.

\emph{Remark 2}.\enspace If in~(\ref{d=0}) the two complex eigenvalues
had been $\lambda+i\eta$ (twice) rather than $\lambda\pm i\eta$ then
the equations would have $D_4\times S^1$ symmetry and~(\ref{d=0})
would already be in normal form: the terms $(\bar\ap\am^2,
\bar\am\ap^2)$ could not be removed.  Compare~\cite{swift88a},
Sect.3.1.

\emph{Remark 3}.\enspace We have been working with Taylor series
truncations at order 3. In fact, normal form theory would allow us to
transform away all those higher order terms in a full Taylor series
extension of~(\ref{d=0}) that did not have $S^1\times O(2)$ symmetry.
However, the higher the order the smaller the likely region of
validity, and there is no guarantee that the system itself exhibits
strict $S^1\times O(2)$ symmetry.  This does not affect robust
(structurally stable) features of the bifurcation scenario, but can
affect sensitive features such as heteroclinic connections and
families of orbits on an invariant torus.  See~\cite{swift88a} for a
careful discussion of this issue for a system closely related to ours.

\smallskip

The advantage of using the variables $\bb$ is that we can now recast
the problem of connecting the solutions for $\delta=0$ to those for
$\delta \ne 0$ as the problem of breaking the $O(2)$ symmetry studied
in~\cite{dangelmayr91a}.  There it is shown that the fully symmetric
system has a standing wave solution $v = \bar w$ and two travelling
wave solutions $(v,0)$ and $(0,w)$.  We indicate these standing and
travelling wave solutions with the symbols $SW^*$ and $TW^*$
respectively.  Translating in terms of $\alpha_+$ and $\alpha_-$ by
identifying $(v,w)$ with $(\bp,\bm)$ and noting that
\begin{equation}
 \ap=0\ \Leftrightarrow\ \bp=0, \quad
 \am=0\ \Leftrightarrow\ \bm=0, \quad
 \bar\ap=\am\ \Leftrightarrow\ \bar\bp=\bm \, ,
\end{equation}
we thus recover the results of Theorem~\ref{bifs} together with the
additional information:

\begin{corollary}
The transformation $\ab\mapsto\bb$ converts the periodic solutions
with $Z_4$ spatio-temporal symmetry into rotationally symmetric ones,
at least up to third order.
\end{corollary}

In particular the Hopf solution $W_\varphi$ (denoted by $H$ in
Figure~\ref{fig:BifDiag}) corresponds to $SW^*$, while the two
solutions $V_{\pm}$ ($SW$ in Figure~\ref{fig:BifDiag}) correspond to
the two solutions $TW^*$.

\begin{figure}
 \centerline{\epsfig{file={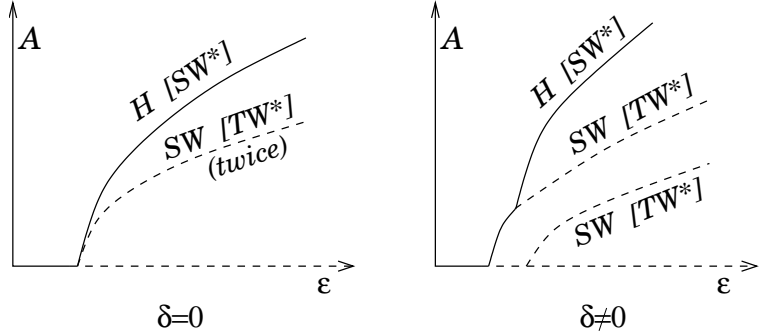},width=100mm}}
 \caption{\label{fig:O2_and_delta} \it Changes in the primary
 bifurcation point as $\delta$ is increased from zero.   The letters
 outside the brackets identify the branches with those in
 Figure~\ref{fig:BifDiag} and refer to the equations in terms of
 $\alpha_{\pm}$.    The letters in square brackets give the 
 names of the branches as viewed in terms of the variables $(v,w)$.}
\end{figure}

Furthermore, it is shown in~\cite{dangelmayr91a} that as the real part of
the symmetry breaking parameter $e$ in equations~(\ref{dangeqs}) moves
away from zero the two $TW^*$ branches separate and the
$SW^*$ solution appears as a secondary bifurcation from
one of the $TW^*$ branches.  In terms of the variables
$\ab$ this implies that as $\delta$ moves away from zero the two $SW$
branches have different starting points, one at $\varepsilon =
\varepsilon_1$ the other at $\varepsilon = \varepsilon_6$.  Moreover
the Hopf branch becomes a secondary branch off the first $SW$ branch.
This is illustrated in Figure~\ref{fig:O2_and_delta}.

\begin{figure}
 \centerline{\epsfig{file={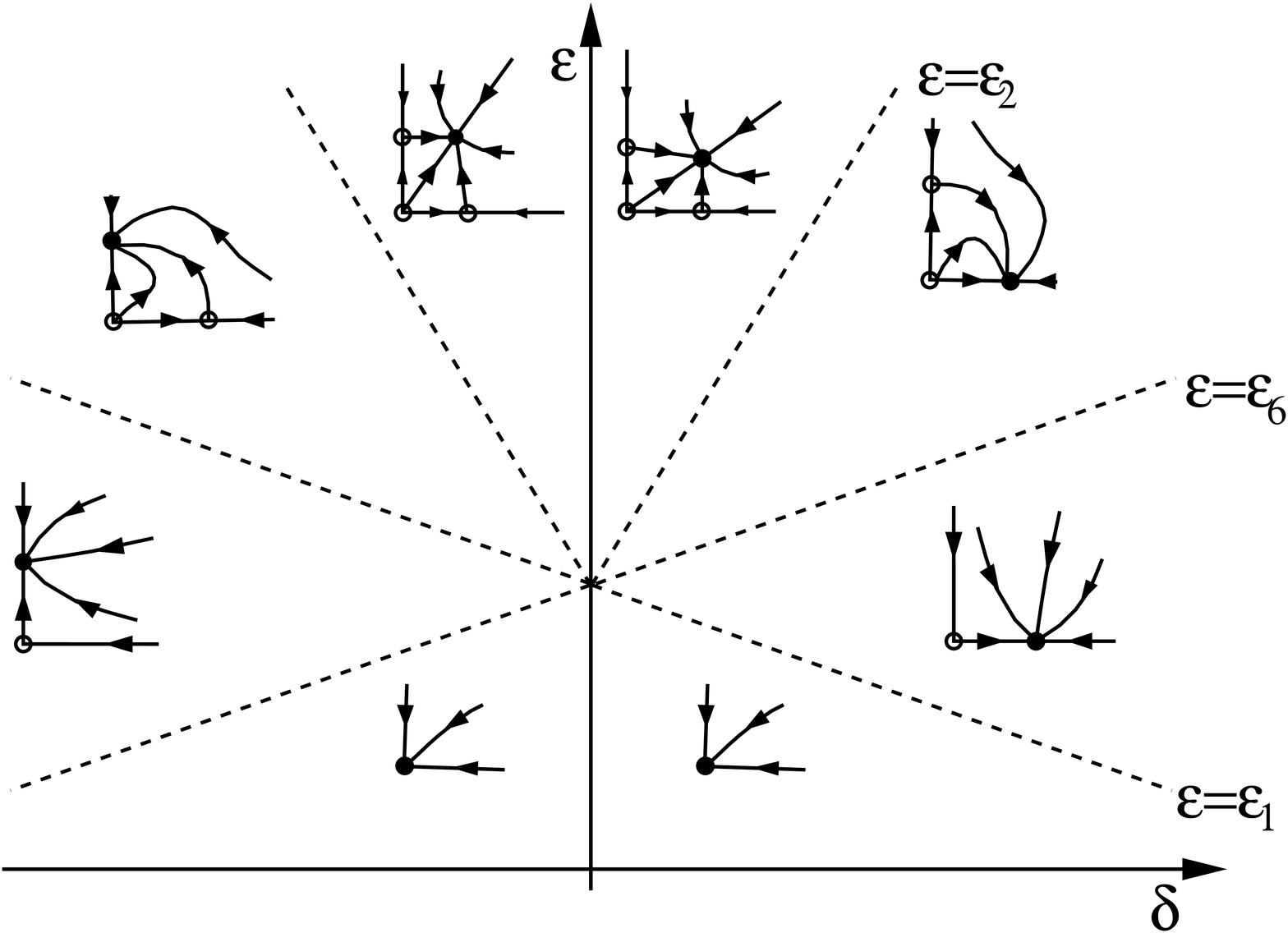},width=100mm}}
 \caption{\label{fig:BifSet} \it Bifurcation set in the neighbourhood
 of $(\eps,\delta)=(\mu^2\eta^2,0)$ in terms of the variables
 $\beta_{\pm}$ (adapted from~\cite{guckenheimer1983}, Figure
 7.5.2 case 1a).   The axes of the small inserts are $|\beta_\pm|$ and
 the filled and open circles indicate stable and unstable fixed points
 respectively.}
\end{figure}

This picture is confirmed by standard bifurcation analysis for systems
with two purely imaginary
eigenvalues~\cite{takens74a,guckenheimer1983}.  Using the $\bb$
coordinates for both $\delta$ zero and nonzero (but small) the
parameters $\eps,\delta$ provide versal unfolding parameters about the
organising centre $(\eps,\delta)=(\mu^2\eta^2,0)$.  In particular, we
can see from Figure 7.5.2 of~\cite{guckenheimer1983} (redrawn
here in terms of $\varepsilon$ and $\delta$ as
Figure~\ref{fig:BifSet}) how the branching of the solutions
$W_{\varphi}$ and $V_{\pm}$ that is simultaneous when $\delta=0$
changes when $\delta\ne0\,$: a succession of two primary bifurcations
with either $\ap=0$ or $\am=0$ is followed by a secondary (Hopf)
bifurcation to a torus of periodic solutions.  From~(\ref{W_phi}) we
also see that the period of these solutions is approximately given by
$2 \pi/\eta$.

To conclude the analysis of the bifurcation diagram for $\delta=0$ we
note that the limit points of the $TW'$ branches of panel A of
Figure~\ref{fig:BifDiag} exist for all sufficiently small $|\delta|$
(including $\delta=0\,$).  Exactly as in the case $\delta \ne 0$ they
are created at a symmetric pair of saddle-node bifurcations of
periodic orbits that annihilate the torus (blue sky).  These branches,
found by direct calculation as in~\cite{dangelmayr91a}, are not
connected to either of the $\ap=0\,,\am=0$ branches when $\delta=0$
and so are not detectable by local bifurcation analysis in that case.

\subsection{Numerical verification}

In order to illustrate the range of applicability of the normal forms
derived in this article we have integrated numerically the
equations~(\ref{fpqeqn}-\ref{Neqn}) that correspond to the cavity in
Figure~\ref{fig:laser}.  The empty cavity modes can be expressed in
terms of Hermite polynomials and are given
by~\cite{siegman1986,ananev1992}:
\begin{equation}
 A_{p q}(x,y,z) = \sqrt{\frac{2^{1-p-q}}{\pi w_x(z) w_y(z) \, p! \, q!}} \, 
                  H_p \left [ \frac{\sqrt{2} x}{w_x(z)} \right ]
                  H_q \left [ \frac{\sqrt{2} y}{w_y(z)} \right ]
                   e^{-x^2/w_x^2(z)} e^{-y^2/w_y^2(z)} 
 \label{GH_modes}
\end{equation}
The indices $p$ and $q$ are nonnegative integers: $p$ is the number of
zeros of the field along the $x$-axis whilst $q$ is the number of
zeros along the $y$-axis. The functions $w_x(z)$ and $w_y(z)$ are $z$
dependent scaling factors that represent the field beam waist,
\emph{i.e.} the spot size radius, along the $x$ and $y$ axis.  They are
functions respectively of $R_x(\theta)$ and $R_y(\theta)$ and are
therefore different for all $\theta \ne 0$.  For each value of $z$ the
cavity modes form an orthonormal basis for the space ${\mathcal H}$.

\begin{figure}
 \centerline{\epsfig{file={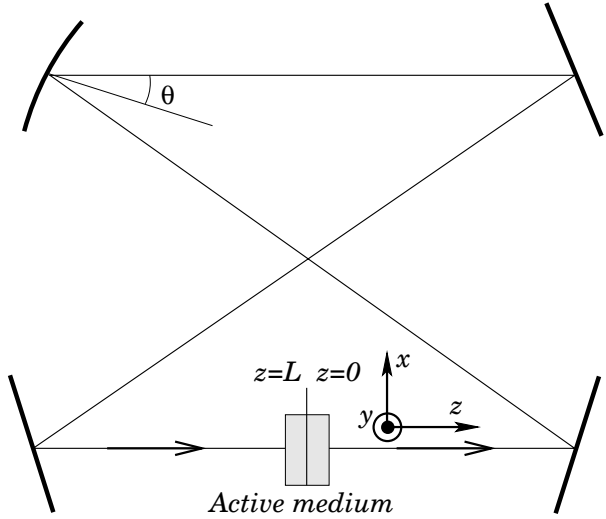},width=80mm}}
 \caption{\label{fig:laser} \it Schematic diagram of a ring cavity
 laser.  The $z$ coordinate is along the cavity axis of length $L$ and
 the points $z=0$ and $z=L$ coincide.}
\end{figure}

We have assumed that there are only two active modes, $(p,q)=\{(1,0),
\, (0,1)\}$.  The first mode is a ``cosine'' mode, \emph{i.e.} its
dependence on the polar angle $\varphi$ is $\cos(\varphi)$, the second
is a ``sine'' mode.  In order to associate the solutions discussed in
the previous section to laser patterns one should keep in mind that
the intensity profiles of the cosine and sine mode consists of two
bright spots aligned horizontally and vertically respectively.

The hypothesis that only two modes are active may not be realistic if
the laser is operated well above threshold, but, on the other hand,
the purpose of these simulations is to illustrate the application of
the normal forms: these are only valid relatively close to threshold
as they have been derived under the assumption that the field
amplitudes are small.

We have scaled the longitudinal coordinate $z$ and the radius of
curvature of the mirror $R$ to the cavity length $L$.   The transverse
dimensions $x$ and $y$ and the beam waists $w_x(z)$ and $w_y(z)$
are scaled to $\sqrt{L \lambda/\pi}$, where $\lambda$ is the laser
wave length.   In these units the beam waists in the plane $z=0$ are
given by~\cite{ananev1992}: 
\begin{equation}
 w_x \equiv w_x(0) = 
  \left [ \frac{R \cos\theta}{2} - \frac{1}{4} \right ]^{1/4} ,
 \quad
 w_y \equiv w_y(0) = 
  \left [ \frac{R}{2 \cos\theta} - \frac{1}{4} \right ]^{1/4} .
\end{equation}
The phase shift per cavity round-trip of the mode $(p,q)$ is given by 
\begin{equation}
 \omega_{p,q} = \left ( p + \frac{1}{2} \right ) \omega_x +
                \left ( q + \frac{1}{2} \right ) \omega_y ,
\end{equation}
where 
\begin{equation}
 \omega_x = \arccos \left [ 1 - \frac{1}{R \cos\theta} \right ] ,
 \quad
 \omega_y = \arccos \left [ 1 - \frac{\cos\theta}{R} \right ] ,
\end{equation}
so that the symmetry breaking parameter $\eta$ used
in~(\ref{betaExample}) is given by
\begin{equation}
 \eta = \frac{|\omega_x - \omega_y|}{ 2 (1 - {\mathcal R})} ,
\end{equation}
where ${\mathcal R}$ represents the total field reflectance of the cavity
mirrors.    The field decay rate $\kappa$ is related to the mirror
reflectance and the cavity round trip time $T_c$ by
\begin{equation}
 \kappa = \frac{1 -{\mathcal R}}{T_c} .
\end{equation}
Finally, the scaling factor ${\mathcal M}$ defined in~(\ref{defcalM}) can
be obtained by writing in polar coordinates in the plane $z=0$ the
cavity modes $A_{p q}$ given by equation~(\ref{GH_modes}) assuming
that $w_x = w_y$ (\emph{i.e.} that the symmetry breaking is small enough for
its effects to be neglected in the coefficients of the nonlinear
terms).  If we separate the radial and angular part and integrate them
separately we obtain that
\begin{equation}
 {\mathcal M} = \frac{2^6}{\pi^2 w_x^4} 
   \int_0^\infty r^5 e^{-2 r^2/w_x^2} \, d r = \frac{1}{\pi^2 w_x^2}\, .
\end{equation}
We have chosen the values of the laser parameters in order to maximise
the speed of integration: $\gamma=1$, ${\mathcal R} = 0.97$, $T_c=0.03$
(hence $\kappa=1$).  Finally, the radius of curvature of the
mirror was set to $R=3/2$ and the angle $\theta$ to $\pi/32$ so that
the symmetry breaking parameter is relatively small, $\eta=0.11377$.
We have simulated the case of one mode being in resonance with the
atomic medium by selecting a value of the detuning equal to the mode
frequency, that is $\delta=\eta\,$.

\begin{figure}
 \centerline{\epsfig{file={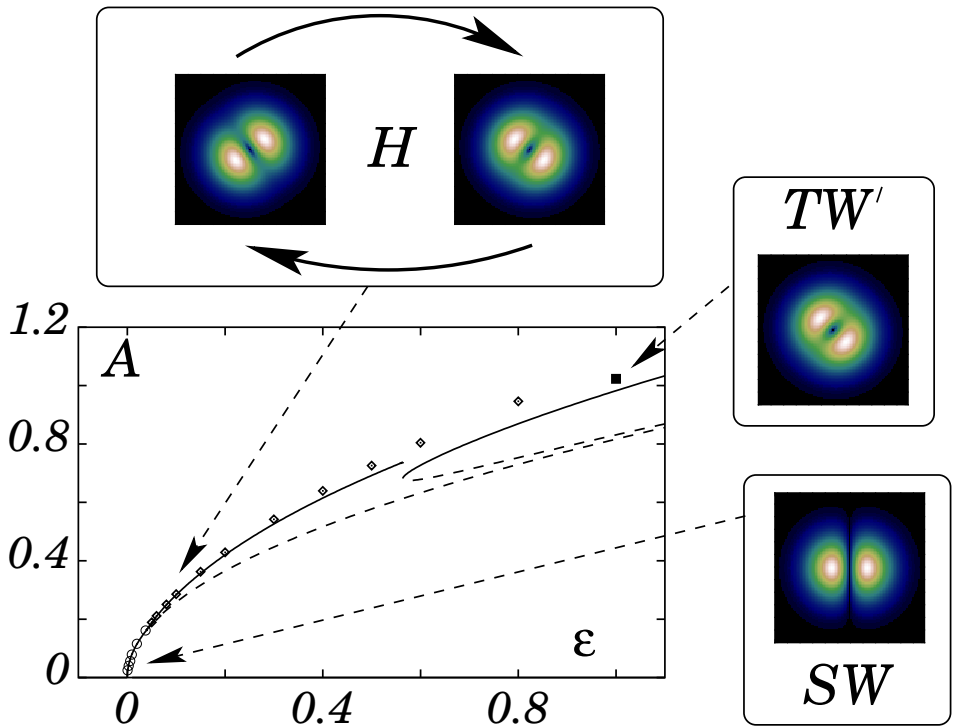},width=130mm}}
 \caption{\label{fig:NumSim} \it Bifurcation diagram of
 equations~(\ref{alphaPlus},~\ref{alphaMinus}) for $\kappa=1$ and
 $\eta=0.11377$, $\delta=\eta$.  The circles, diamonds and squares
 represent the results of numerical integration of the
 equations~(\ref{fpqeqn}--\ref{Neqn}) for the cavity in
 Figure~\ref{fig:laser} using the parameters detailed in the text.
 The circles correspond to stationary single mode solutions ($SW$),
 the diamonds to periodic solutions ($H$) and the square represents a
 stationary two mode solution ($TW'$).  The images in the inserts show
 the stationary field in the case of the $SW$ and $TW'$ solution and
 the two oscillating patterns in the case of the periodic solution.  }
\end{figure}

The program to integrate equations~(\ref{fpqeqn}-\ref{Neqn})
represents the polarisation and population inversion on a rectangular
grid thus transforming the partial differential equation into ordinary
differential equations for the values of the fields at the grid
points.  These are integrated using a variable order variable step
method~\cite{ode}.  The projection integral in equation~(\ref{fpqeqn})
is computed using Gaussian quadrature~\cite{press1992}.

We have run a set of simulations for different values of the pump
parameter $\chi = \kappa( 1+ \varepsilon)$.  The results of these
simulations are shown in Figure~\ref{fig:NumSim} overlaid on the
bifurcation diagram of equations~(\ref{alphaPlus},~\ref{alphaMinus}).
For small values of $\varepsilon$ the agreement between the full
equations and the predictions of the normal form is excellent.  The
laser switches on in a single mode stationary solution (see inset $SW$
of Figure~\ref{fig:NumSim}).  This becomes unstable for larger pump
values and a periodic orbit appears that oscillates between two
patterns (see inset $H$ of Figure~\ref{fig:NumSim}).  As the smallness
parameter $\varepsilon$ becomes larger the agreement between the
normal forms and the full equations becomes worse.  In both cases the
periodic orbit disappears and is replaced by a stationary two mode
solution (panel $TW'$ in Figure~\ref{fig:NumSim}), but the value of
the pump parameter at which this happens is much larger for the full
equations than for the normal forms.  However, it is quite unlikely
that only two modes of the laser would be active for these values of
the pump parameter and is therefore debatable whether one should rely
on the normal forms as a guide to the laser behaviour in this region
of parameter space.  Be that as it may, it is clear from the numerical
simulations that the normal forms are an excellent vehicle for
understanding the dynamics of a laser with broken symmetry both on a
qualitative and quantitative level for values of the smallness
parameter up to roughly $1/2$.

\section{Conclusions}

Normal forms have long been known as an extremely useful tool for
studying the bifurcation structure of generic models derived solely on
the basis of symmetry considerations.  However, the connection between
the parameters of a specific model and the coefficients of the
corresponding normal forms is seldom made explicit in the bifurcation
literature.  In this paper we have addressed this issue for the case
of a two level Maxwell-Bloch laser and have derived the formulae that
express the coefficients of the normal forms in terms of the
parameters of the laser.  This is important for two reasons.  On the
one hand, we can identify the regions where the predictions derived
from the normal forms are in quantitative agreement with the
predictions of the full model in contrast to those where the agreement
is only qualitative, as illustrated in Figure~\ref{fig:NumSim}.  On
the other hand, we can clearly disentangle the ``model-independent''
features of a theory from the ``model-dependent'' features.
Model-independent features are, for instance, the type of solutions
and the bifurcations observed, which depend on the symmetry and on the
first bifurcation and are, therefore, common to all models having the
same symmetry and the same first bifurcation. Model-dependent features
include the positions of the bifurcations and the widths of the
stability windows in the control parameter space, and depend on
important physical aspects of each individual model that do not affect
the symmetry. In optics, for instance, the number of energy levels
included in the description of media interacting with light does not
change the symmetry of the model. When more than one model is
available, the ability to separate model-dependent and
model-independent features is vital for ascertaining which model is
the most accurate.  In principle, this could be done by systematically
comparing experiments and models over a wide range of control
parameters.  In practice, however, such checks are very difficult to
carry out without a knowledge of where to look in parameter space for
model-dependent features, since outside these regions all models with
the same symmetry behave in much the same way.

In addition we have combined in this paper the analysis of two key
bifurcation phenomena that apply to the Maxwell-Bloch laser and
exhibited the relation between them in parameter space.  Each of these
phenomena has been well-studied in a variety of theoretical and
applied contexts in the literature, but we are not aware of other
studies that incorporate the full picture as we have presented it
here.  The curious interplay of travelling waves and standing waves is
a new observation.    

The role played by the symmetry generated by $\tau$ at $\delta=0$ in
organizing the bifurcations, as shown in Figure~\ref{fig:O2_and_delta}
and~\ref{fig:BifSet}, is an important new result of
section~\ref{TwoModes}. From a physical point of view, this additional
symmetry is present when the maximum of the gain profile $\omega_A$ is
between two cavity modes and these modes have the same gain.  Up to
the second order expansion of the gain line in small detunings, this
happens for all active media. For large detunings, the gain profile
needs to be symmetrical with respect to $\omega_A$, a condition
verified in most lasers. We can conclude then that this symmetry is
useful in understanding the interaction of two laser modes for a class
of laser much larger than that described by the Maxwell-Bloch model
used in this paper.

A final point worth remarking is that our derivation of the normal
forms takes place in a broad context of breaking rotational symmetry.
In this paper we have analysed in detail the case of an astigmatic
laser, but the same normal forms could be used to study the symmetry
breaking induced for example by the interaction of a metallic wave
guide and an aperture, the case studied in~\cite{papoff99a,papoff99b}.

\noindent \textbf{Acknowledgement}

\noindent 
We thank Dr Philip Aston of the University of Surrey and Dr Alastair
Rucklidge of the University of Leeds for several valuable discussions
on symmetry-breaking issues.

\begin{appendix}

\section{Detailed calculations of the centre manifold expansion}
\label{DetailedCalc}

At the heart of the centre manifold reduction~\cite{carr81a} lies the
splitting~(\ref{SplittingU}) of the fields $\bu\,$.  Let $\bu_{\|}$
denote the complex linear subspace of the function space $\mathcal H$
generated by the active modes $\bu_n$, $n \in {\mathcal J}$.  Centre
manifold theory ensures that the local dynamics of
equations~(\ref{fpqeqn}-\ref{Neqn}) can be obtained by considering the
restriction of the laser equations to the centre manifold of the
trivial solution at bifurcation.  On this manifold we can consider the
non-active modes as functions of the active modes, that is $\bu_\perp
= \bu_\perp (\bu_{\|})$.  The expression for the centre manifold is
obtained by substituting this functional relation into the laser
equations~(\ref{fpqeqn}-\ref{Neqn}).  However, this is quite
intractable unless we make some further simplifications.  We therefore
assume that we are sufficiently close to threshold so that the
amplitudes of the active modes are small and we can approximate the
centre manifold by the leading terms of its power series expansion in
the amplitudes of the active modes.

In order to understand which terms of this series we need to compute
it is convenient to write out in full the
equation~(\ref{EqActModesFull}) for the amplitudes of the active
modes:
\begin{eqnarray}
 & & \displaystyle \frac{d}{d t} f_n =
  \frac{1}{{\mathcal N}_n} 
  \left ( 
   \begin{array}{ccc} 1 + \lambda_n & 1 & 0 \end{array}
  \right ) \times \label{ActModEqn1} \\
 & & 
  \left (
   \begin{array}{l}
    \displaystyle 
     - \sum_{p \in {\mathcal J}} \beta_p  
         \frac{( 1 + \lambda_p )}{{\mathcal N}_p} (A_n, A_p) f_p
     - (A_n, F_\perp) +
     \chi \sum_{p \in {\mathcal J}} \frac{(A_n, A_p)}{{\mathcal N}_p} f_p +
     (A_n, P_\perp) \\*[4mm]
    \displaystyle
     -\chi \sum_{p \in {\mathcal J}} \frac{(A_n, A_p)}{{\mathcal N}_p} f_p -
     (A_n, P_\perp) + \\*[4mm]
     \hspace{30mm} \displaystyle
      \sum_{p \in {\mathcal J}}  
         \frac{( 1 + \lambda_p )}{{\mathcal N}_p} (A_n, (\chi + N) A_p) f_p +
     (A_n, (\chi + N) F_\perp) \\*[4mm]
     -\gamma \displaystyle \left [ N - \frac{1}{2} 
       (F \bar P + \bar F P) \right ]
   \end{array}
  \right ) \nonumber
\end{eqnarray}
where the symbol $(f,g)$ represents the $L_2$ inner product of two
functions $f(x,y)$ and $g(x,y)\,$.  Using the orthogonality
relation~(\ref{OrthoRelUPerp}), we can eliminate the terms linear in
$F_\perp$ and $P_\perp$ in equation~(\ref{ActModEqn1}) and write it as
\begin{eqnarray}
 \displaystyle \frac{d}{d t} f_n & = & 
  \frac{1}{{\mathcal N}_n^2} 
   \left [ \chi \left ( 1 + 2 \lambda_n \right ) -
          \beta_n \left ( 1 + \lambda_n \right )^2
  \right ] f_n \nonumber \\
 & & \hspace{10mm}
  - \frac{1}{{\mathcal N}_n} \sum_{p \in {\mathcal J}}
        \frac{(1+\lambda_p)}{{\mathcal N}_p} (A_n, N A_p ) f_p +
  \frac{1}{{\mathcal N}_n} 
        (A_n, N F_\perp ) \nonumber \\ 
 & =  & \lambda_n f_n -  
  \frac{1}{{\mathcal N}_n} \sum_{p \in {\mathcal J}}
        \frac{(1+\lambda_p)}{{\mathcal N}_p} (A_n, N A_p ) f_p 
  + \frac{1}{{\mathcal N}_n} 
        (A_n, N F_\perp )  ,
   \quad n \in {\mathcal J}\,.   \nonumber \\
 & & \hspace{10mm} \label{ActModEqn2} 
\end{eqnarray}
Inspection of equation~(\ref{Neqn}) for the population inversion shows
that it is at least of second order in the amplitudes of the active
modes.  This holds also for $F_\perp$ since the source of the electric
field is ultimately the term in $N$ in the polarisation
equation~(\ref{Peqn}).  Therefore the last two terms
in~(\ref{ActModEqn2}) are respectively of third and fourth order in
the amplitudes of the active modes.  As we are interested only in a
low order expansion of~(\ref{ActModEqn2}), we can drop the last term
and expand the second term up to third order.

The lowest order terms in the expansion of the nonlinear terms on the
right hand side of~(\ref{Neqn}) are
\begin{eqnarray}
 & & \frac{1}{2}
 \left [ 
  \sum_{j \in {\mathcal J}} \frac{(1+\lambda_j) A_j}{{\mathcal N}_j} f_j
 \right ] \, \,
 \left [ \chi
  \sum_{k \in {\mathcal J}} \frac{\bar A_k}{\bar {\mathcal N}_k} \bar f_k
 \right ] \, \, + \, \, \mbox{c.c.} = \nonumber \\
 & & \hspace{20mm}
  \frac{1}{2} \sum_{j,k \in {\mathcal J}} 
    \frac{(1 + \lambda_j) \chi}{{\mathcal N}_j \bar {\mathcal N}_k} 
    A_j \bar A_k f_j \bar f_k +
  \frac{1}{2} \sum_{j,k \in {\mathcal J}} 
    \frac{(1 + \bar \lambda_j) \chi}{\bar {\mathcal N}_j {\mathcal N}_k} 
    \bar A_j A_k \bar f_j f_k = \nonumber \\
 & & \hspace{20mm}
  \sum_{j,k \in {\mathcal J}} 
    \frac{[1 + (\lambda_j + \bar \lambda_k)/2]}
         {{\mathcal N}_j \bar {\mathcal N}_k} 
    \chi A_j \bar A_k f_j \bar f_k \, . \label{NonLinTermN2nd} 
\end{eqnarray}
This suggests expanding the population inversion as
\begin{equation}
 N = \sum_{j,k \in {\mathcal J}} N_{j k} f_j \bar f_k 
 \label{expandN}
\end{equation}
where the expansion coefficients $N_{j k}$ are such that $N_{j k} =
\bar N_{k j}$ in order to ensure that $N$ is real.  Expressions for
the $N_{jk}$ can be obtained by substituting~(\ref{expandN})
into~(\ref{Neqn}).  Before doing so it is convenient to expand
separately the left hand side of~(\ref{Neqn}), \emph{i.e.} to compute
the time derivative of $N$ in terms of the time derivative of the
amplitudes of the active modes:
\begin{equation}
 \displaystyle  \frac{ \partial N}{\partial t} =
  \sum_{j,k \in {\mathcal J}} N_{j k}
   \left ( f_j \frac{d}{d t} \bar f_k +
           \bar f_k \frac{d}{d t} f_j 
   \right ) .
\end{equation}
Substituting the lowest order (\emph{i.e.} the linear) terms of the
active modes equation (\ref{ActModEqn2}) into this expression we
obtain
\begin{equation}
 \displaystyle  \frac{ \partial N}{\partial t} =
  \sum_{j,k \in {\mathcal J}} 
    (\lambda_j + \bar \lambda_k) N_{j k} f_j \bar f_k .
 \label{TimeDerN2nd}
\end{equation}
Substituting~(\ref{TimeDerN2nd}) and~(\ref{NonLinTermN2nd})
into~(\ref{Neqn}) we obtain
\begin{equation}
 \sum_{j,k \in {\mathcal J}} 
  (\lambda_j + \bar \lambda_k) N_{j k} f_j \bar f_k =
  - \gamma \sum_{j,k \in {\mathcal J}} 
     \left \{ N_{j k} + 
     \frac{[1 + (\lambda_j + \bar \lambda_k)/2]}
          {{\mathcal N}_j \bar {\mathcal N}_k} 
     \chi A_j \bar A_k 
    \right \} f_j \bar f_k\, ,
\end{equation}
from which we finally derive
\begin{equation}
 N_{j k} = - \frac{1 + (\lambda_j + \bar \lambda_k)/2}
                  {[ 1 + (\lambda_j + \bar \lambda_k)/\gamma ] 
                  {\mathcal N}_j {\mathcal N}_k}
             \chi A_j \bar A_k\, , \hspace{15mm} j, k \in {\mathcal J} .
\end{equation}
Substituting this expression into the third term of~(\ref{ActModEqn2})
and keeping in mind that the last term of~(\ref{ActModEqn2}) can be
neglected because it is of higher order we obtain the 
equation~(\ref{ActModEqn}) for the amplitudes of the active modes.

\section{Space dependent pump parameter}
\label{SpaceDepPump}

If the pump parameter is space-dependent the calculations to obtain
the normal forms become much more involved.  The principle behind the
derivation remains unchanged, we still need to separate the slow from
the fast modes, but its actual realisation is a non trivial affair.
If the pump is flat, \emph{i.e.} $\chi$ does not depend on the transverse
coordinates, the cavity modes diagonalise the linear equations:
therefore the slow modes are a subset of the cavity modes and we can
relatively easily split the laser equations into two blocks.  If the
pump parameter is space dependent this is no longer true.  The
eigenmodes $G_n(x,y)$  of the linearised laser equations~(\ref{fpNeqnLin})
are no longer cavity modes, but can be expressed as linear
superpositions of them:
\begin{equation}
 G_n(x,y) = \sum_{k=1}^\infty g_k A_k(x,y) .
 \label{decomposeGn}
\end{equation}
In order to separate the slow and fast dynamics we then need to project
the equations~(\ref{fpqeqn}-\ref{Neqn}) onto the modes $G_n$ using
the decomposition~(\ref{decomposeGn}).  This is of course possible in
principle, but rather hard to do in practice, especially since the modes
$G_n(x,y)$ are usually only known numerically.  Moreover, the laser
equations are not diagonal on the basis of the modes $G_n$  and  the
resulting equations for the amplitudes of these modes contain sums over
all the coefficients $\beta_n$ of equation~(\ref{fpqeqn}).   It is
likely that a symbolic algebra package could be effective in 
splitting fast and slow variables, after which the procedure detailed in
Appendix~\ref{DetailedCalc} can in principle be applied 
to obtain the normal forms.

\end{appendix}

\newpage

\end{document}